\documentclass[pdflatex,sn-chicago]{sn-jnl}

\usepackage{graphicx}%
\usepackage{multirow}%
\usepackage{amsmath,amssymb,amsfonts}%
\usepackage{amsthm}%
\usepackage{mathrsfs}%
\usepackage[title]{appendix}%
\usepackage{xcolor}%
\usepackage{textcomp}%
\usepackage{manyfoot}%
\usepackage{booktabs}%
\usepackage{listings}%

\usepackage{multirow}
\usepackage{enumerate}
\usepackage{url} 
\usepackage{subfiles}
\usepackage{mathtools}
\usepackage[ruled,vlined,linesnumbered]{algorithm2e}
\usepackage{commath}
\usepackage{placeins}
\usepackage[subrefformat=parens,labelformat=parens]{subfig}
\usepackage{xcolor}
\usepackage[font=small]{caption}
\usepackage{enumitem}
\usepackage{soul}
\usepackage{graphics}
\usepackage{moresize}
\usepackage{natbib}

\newcommand\numberthis{\addtocounter{equation}{1}\tag{\theequation}}

\newcommand\particle{\boldsymbol{s}}
\newcommand\parameter{\boldsymbol{\theta}}
\newcommand\hidden{\textbf{X}}
\newcommand\data{\textbf{D}}

\theoremstyle{thmstyleone}%
\theoremstyle{thmstyletwo}%

\theoremstyle{thmstylethree}%

\begin{document}

\title[Generalized Bayesian Multidimensional Scaling and Model Comparison]{Generalized Bayesian Multidimensional Scaling and Model Comparison}


\author[1]{\fnm{Jiarui} \sur{Zhang}}\email{jiarui\_zhang@sfu.ca}

\author[1]{\fnm{Jiguo} \sur{Cao}}\email{jiguo\_cao@sfu.ca}

\author*[1]{\fnm{Liangliang} \sur{Wang}}\email{lwa68@sfu.ca}

\affil*[1]{\orgdiv{Department of Statistics and Actuarial Science}, \orgname{Simon Fraser University}, \orgaddress{\street{8888 University Drive}, \city{Burnaby}, \postcode{V5A 1S6}, \state{BC}, \country{Canada}}}


\abstract{
    Multidimensional scaling (MDS) is a fundamental dimension reduction technique widely applied across diverse fields to represent objects as points in a lower-dimensional space based on pairwise dissimilarities. While classical and scalable non-Bayesian MDS methods are valued for their computational efficiency, they primarily focus on optimization, yielding only point estimates and thus neglecting model uncertainty. This limitation motivates the development of Bayesian MDS (BMDS) frameworks, which offer a probabilistic perspective, model uncertainty, and provide posterior distributions for principled uncertainty quantification. However, existing BMDS methods predominantly assume Euclidean distance and Gaussian noise in the observed dissimilarities, which limits robustness and applicability across diverse domains like text mining. Furthermore, Bayesian inference for these models faces computational challenges when scaling to large datasets. To overcome these issues, we propose a Generalized Bayesian Multidimensional Scaling (GBMDS) framework that incorporates flexible dissimilarity metrics and robust non-Gaussian error structures, prioritizing uncertainty quantification, robustness, and model flexibility. To facilitate efficient inference and robust model selection within this framework, we design an adaptive annealed Sequential Monte Carlo (ASMC) algorithm. This ASMC approach mitigates computational burdens in large-scale applications, leverages existing MCMC proposals for ease of implementation, and provides nearly unbiased estimators of marginal likelihoods, enabling rigorous Bayesian model comparison via Bayes factors. The GBMDS framework thus enhances estimation accuracy and robustness by accommodating complex dissimilarities and heavy-tailed distributions.}




\maketitle

\section{Introduction}
\label{sec:introduction}

Multidimensional scaling (MDS) is a dimension reduction technique that represents objects as points in a lower-dimensional space based on a given set of pairwise dissimilarities between objects. The goal is to preserve the relative distance from the original high-dimensional space.  MDS is widely applied across various fields, such as psychology, social science, and genomics, to uncover hidden patterns, explore relationships in complex datasets, and facilitate visualization. By transforming high-dimensional data into a spatial representation, typically two or three dimensions, MDS enables researchers to detect clusters, trends, and structures that might otherwise be difficult to interpret. It also supports exploratory data analysis by identifying the principal dimensions underlying dissimilarities, making it a valuable tool for both data visualization and statistical inference.

MDS methods are broadly categorized into metric and non-metric approaches. Metric MDS assumes that dissimilarities represent numerical distances and satisfy metric properties such as symmetry and the triangle inequality. It is particularly appropriate when the underlying data structure follows Euclidean geometry or when a specific distance function, such as the Euclidean or Mahalanobis distance, is justified. Non-metric MDS, on the other hand, is designed for ordinal or rank-based dissimilarities, where the absolute values of distances are not meaningful, and only the relative ordering of dissimilarities matters. This method is widely used in applications where dissimilarities do not adhere to strict metric properties, such as in psychological or perceptual studies. Despite differences in assumptions, both MDS approaches aim to embed high-dimensional data into a lower-dimensional space while preserving the original similarity structure as faithfully as possible. Our study primarily focuses on metric MDS. For a comprehensive review of modern MDS methods, see \cite{borg2005modern}.

Classical multidimensional scaling (CMDS), introduced by \citet{torgerson1952multidimensional}, is an essential method for metric MDS that reconstructs a low-dimensional Euclidean representation from a given dissimilarity matrix. CMDS is computationally efficient for small datasets; however, its scalability is limited by the extensive memory and time requirements associated with storing and processing the full dissimilarity matrix for large sample sizes. To address this, scalable alternatives have been developed. For example, landmark MDS algorithm \citep{de2004sparse} first applies CMDS to a representative subset of data (landmark points) and then projects the remaining data onto the resulting low-dimensional configuration using a distance-based triangulation procedure. Another popular approach includes Divide-and-Conquer MDS \citep{delicado2025multidimensional}, which preserves local structure by splitting the dataset into overlapping patches, performing independent MDS on each, and then aligning and combining the partial configurations via Procrustes transformation. For a comprehensive review of the scalable non-Bayesian MDS methods, see \citet{delicado2025multidimensional}.

While classical and scalable non-Bayesian MDS methods are valued for their computational efficiency, they primarily focus on optimization and yield only point estimates. This limitation motivates the development of Bayesian MDS frameworks, which offer a probabilistic perspective on dimension reduction. The Bayesian approach explicitly models uncertainty and noise in the observed dissimilarities and provides posterior distributions over the latent coordinates, allowing for principled uncertainty quantification and greater flexibility through the incorporation of prior structures. Given these methodological advantages, and since we prioritize uncertainty quantification, robustness, and model flexibility over scalability, our paper will focus on Bayesian formulations of MDS, leaving comparisons with large-scale non-Bayesian MDS methods beyond the scope of this study.

\citet{oh2001bayesian} introduced a Bayesian multidimensional scaling approach (BMDS) that models observed dissimilarities as noisy observations of underlying Euclidean distances, incorporating measurement errors explicitly within a probabilistic model. Object locations in the low-dimensional space are estimated using Markov chain Monte Carlo (MCMC) sampling. Their results indicate that BMDS provides more accurate embeddings than CMDS, particularly when dissimilarities contain substantial measurement errors, the Euclidean assumption is violated, or the latent dimension is misspecified.

 The Bayesian approach to the MDS problem has gained increasing attention due to its ability to incorporate prior information and handle measurement uncertainty. Several extensions of BMDS have been proposed to enhance its applicability. \citet{oh2007model} integrated Bayesian model-based clustering within the BMDS framework in \citet{oh2001bayesian} to facilitate clustering in high-dimensional spaces. \citet{bakker2013bayesian} modeled observed distances using a log-normal distribution and estimated posteriors via squared error loss minimization. \citet{lin2019bayesian} employed a $t$-distribution for object locations, improving robustness against outliers and enabling variable selection through a latent multivariate regression structure. \citet{liu2024bayesian} extended BMDS to hyperbolic manifolds, where dissimilarities are modeled as hyperbolic distances. Advancements in sampling methods have further improved the efficiency of BMDS. \citet{gronau2020bayesian}  introduced differential evolution MCMC to enhance sampling performance, particularly in the context of psychologically interpretable metrics such as Euclidean and Minkowski distances. 
  \citet{holbrook2020massive}  leveraged Hamiltonian Monte Carlo (HMC)  \citep{neal2011mcmc} and massive parallelization using multi-core CPUs and GPUs to accelerate inference in BMDS, with applications in phylogenetics. Despite these developments, a comprehensive Bayesian framework that accommodates non-Gaussian errors and extends beyond Euclidean spaces remains an open challenge. Furthermore, traditional MCMC-based inference often struggles with the multimodal posterior landscapes common in MDS, limiting its scalability and efficiency.

The existing BMDS methods have several limitations. First, most BMDS methods assume Euclidean distance as the primary dissimilarity metric. However, Euclidean distance may not be suitable for certain applications. In medical imaging and 3D face recognition, surface alignment and shape comparison often rely on the Gromov-Hausdorff distance \citep{memoli2011gromov} or the partial embedding distance \citep{bronstein2006generalized}, which capture geometric distortions more effectively than Euclidean metrics. In text mining, Cosine dissimilarity \citep{li2013distance} is preferred, as it measures the angle between document vectors and remains invariant to document length.
Similarly,  Jaccard dissimilarity \citep{jaccard1901etude} is widely used for set-based comparisons, particularly in short-text clustering. The lack of BMDS frameworks accommodating these alternative dissimilarities limits its applicability across diverse fields. Second, existing BMDS methods predominantly assume Gaussian noise in the observed dissimilarities, resulting in a lack of robustness and generality. Real-world datasets often exhibit heavy-tailed or skewed error distributions, leading to biased or inefficient estimates when using Gaussian-based models. Developing BMDS methods that incorporate more flexible error structures, such as Student’s $t$-distributed errors or nonparametric approaches, remains an open problem. Third, research on model comparison within the BMDS framework remains scarce, particularly concerning the impact of different dissimilarity distributions. Most existing work focuses on comparing Bayesian and frequentist MDS solutions in specific application domains. A systematic evaluation of BMDS models incorporating diverse dissimilarity metrics and error structures would provide deeper insights into their relative performance. Fourth, 
the increasing availability of high-dimensional data presents computational challenges for BMDS. Bayesian inference, while offering flexibility and uncertainty quantification, is often computationally expensive. Scaling BMDS methods to handle large datasets efficiently remains a critical issue, necessitating advances in inference techniques such as variational approximations, parallelized MCMC, or SMC methods.

Sequential Monte Carlo (SMC) methods have become an important alternative method for conducting Bayesian inference \cite[see][for an introduction to SMC]{doucet2001sequential, doucet2009tutorial, chopin2020introduction}. In general, SMC employs a population of weighted particles that evolve sequentially, allowing for greater parallelization and adaptability. SMC methods approximate a sequence of probability distributions using importance sampling and resampling mechanisms, yielding a flexible framework for unbiased estimation of marginal likelihoods \citep{doucet2006efficient}. A particularly relevant variant of SMC is the SMC sampler \citep{del2006sequential, dai2022invitation}, also known as annealed SMC (ASMC) \citep{wang2021adaptive}. ASMC leverages an annealing schedule, similar to simulated tempering, to smoothly transition from an easy-to-sample distribution to the target posterior. This gradual adaptation reduces the risk of getting trapped in local modes and improves exploration of complex posterior landscapes under the same computational budget as standard MCMC. 
Notably, ASMC yields unbiased marginal likelihood estimators when using a fixed sequence of annealing parameters and nearly unbiased estimators when employing an adaptive annealing scheme. The estimated marginal likelihood enables robust Bayesian model comparison via Bayes factors \citep{jeffreys1935some, han2001markov, zhou2016toward, wang2020annealed}. The efficiency of ASMC has been demonstrated in diverse applications, including epidemiology \citep{del2012adaptive}, phylogenetics \citep{wang2020annealed}, and solving nonlinear differential equation systems \citep{wang2021adaptive}. For a comprehensive review of recent advances in ASMC, see \citet{dai2022invitation}.

To address key limitations in existing BMDS methods, such as reliance on Euclidean metrics, Gaussian error assumptions, and computational inefficiencies, we propose a generalized Bayesian multidimensional scaling (GBMDS) framework. GBMDS extends BMDS by incorporating flexible dissimilarity metrics and non-Gaussian error structures, enhancing robustness and applicability. We design an adaptive inference framework based on the annealed SMC algorithm to obtain Bayesian inference for the proposed GBMDS model. Unlike standard SMC methods that require designing novel proposal distributions, our approach leverages existing Metropolis-Hastings proposals. This facilitates seamless integration with established MCMC methodologies, reducing implementation complexity. Our adaptive annealed SMC algorithm is designed to handle dynamically increasing data, parameter dimensionality, and hidden variables. It enables adaptive inference to allow real-time updates as new data become available. This framework enables practitioners to update and refine posterior estimates incrementally as new data arrive, making it particularly suited for streaming or large-scale datasets. To enhance scalability, the proposed adaptive scheme processes large datasets in smaller batches, enabling efficient sequential Bayesian updates while mitigating computational bottlenecks.

Our contributions can be summarized as follows: (\romannumeral 1) We generalize BMDS to accommodate non-Gaussian errors in the pairwise dissimilarities. This allows our model to handle data with heavy-tailed or skewed distributions, improving robustness and accuracy. (\romannumeral 2) Unlike conventional BMDS models restricted to Euclidean distances, GBMDS supports a broad class of dissimilarity measures, making it applicable to diverse domains, such as text mining. (\romannumeral 3) 
We introduce an efficient adaptive Bayesian inference framework based on annealed SMC. This approach mitigates computational burdens in large-scale applications while leveraging existing Metropolis-Hastings proposals, ensuring ease of implementation. (\romannumeral 4) Our framework provides nearly unbiased estimators of marginal likelihoods as a byproduct of sampling, which facilitates model selection through Bayes factors. This enables rigorous comparison between competing BMDS models with different error structures, dissimilarity metrics, and dimensions. (\romannumeral 5) We evaluate the proposed GBMDS framework in three simulation studies and four real-world applications, demonstrating superior estimation accuracy and robustness across diverse dissimilarity metrics compared to benchmark methods. The implementation of our proposed method and applications are available at \url{https://github.com/SFU-Stat-ML/GBMDS}. The implementation of the simulation studies is available at \url{https://github.com/nunujiarui/GBMDS_simulation}.

The rest of this article is organized as follows. Section \ref{sec:model} describes the models for BMDS, including our proposed GBMDS model, specifications, model comparison, and identifiability considerations (Section \ref{sec:model_GBMDS}–\ref{sec:identifiability}). Section \ref{sec:GBMDS_methodology} depicts the implementation for the GBMDS model: Sections \ref{sec:ASMC_initialization} and \ref{sec:ASMC_main} detail the initialization and inference procedure, while Section \ref{sec:ASMC_adaptive} outlines the adaptive mechanism with the annealed SMC algorithm. Simulations and real-world examples are presented in Sections \ref{sec:simulations} and \ref{sec:examples}. Finally, the conclusion and discussion are provided in Section \ref{sec:conclusion}.

\section{BMDS Models}
\label{sec:model}

Suppose we have a set of $n$ objects in the study. Let $\textbf{Z} = \left\{ \mathbf{z}_1, \ldots, \mathbf{z}_n\right\}$ be a set of observed points with $\mathbf{z}_i = (z_{i,1}, \ldots, z_{i,q})^{\top} \in \mathbb{R}^{q}$ representing the values of $q$ attributes in object $i$. The value of $q$ is usually high, which makes the visualization of the points in their original dimension hard. Let $\textbf{D}$ be the matrix of dissimilarities with entry $d_{i,j}$ as the dissimilarity between objects $i$ and $j$. The dissimilarity matrix $\textbf{D}$ is computed from the observed data $\mathbf{z}_1, \ldots, \mathbf{z}_n$ with specific dissimilarity metrics such as the Euclidean metric. Dissimilarity metrics used in this study will be detailed in Section \ref{sec:model_GBMDS}. A formal definition of the metric space is given in \textit{Supplementary}.

Let $\mathbf{x}_i = (x_{i,1}, \ldots, x_{i,p})^{\top} \in \mathbb{R}^{p}$ be the unobserved vector representing the values of $p$ significant attributes in object $i$. The goal of MDS methods is to find the set of points $\textbf{X} = \left\{ \mathbf{x}_1, \ldots, \mathbf{x}_n\right\}$ such that $d_{i,j}$ and  $\| \mathbf{x}_i - \mathbf{x}_j \|_{p}$ are as close as possible, where $\|\cdot \|_{p}$ represents the $L^p$ norm. In such a manner, the given dissimilarities are well-reproduced by the resulting configuration. We refer to this process as object configuration \citep{oh2001bayesian}, which describes the estimation of values for objects' significant attributes.

CMDS is a commonly used dimension reduction technique for metric MDS developed by \citet{torgerson1952multidimensional}. CMDS assumes the dissimilarity to be Euclidean and takes the pairwise dissimilarities as inputs and outputs the coordinates of points in a low-dimensional space up to locations, rotations and reflections. Numerical optimization techniques can be used to find a solution to the minimization problem below:
\begin{equation}\label{metric_MDS}
\text{min} \sum_{\substack{i, j = 1 \\ i \neq j}}^n \left( d_{i,j} - 
\|  \mathbf{x}_i - \mathbf{x}_j \|_{p} \right)^{2}.
\end{equation}

The minimizers can be expressed analytically in terms of matrix eigendecompositions when the input dissimilarities satisfy the metric inequality and can be represented by Euclidean distances. CMDS can retrieve the complete configuration of objects (up to location shift) when the dissimilarities are precisely equal to the distances in the low-dimensional space and the dimension is appropriately specified. However, the dissimilarities between observed points are usually contaminated by errors, and the underlying dimensions are often unknown.

\subsection{Generalized Bayesian multidimensional scaling}
\label{sec:model_GBMDS}

While Euclidean distance is one of the most widely used distance measures, it is not scale-invariant, meaning that distances computed from features might be skewed depending on the units. Moreover, Euclidean distance becomes less useful as the dimensionality of the data increases. To satisfy the various needs of different tasks, we develop a general framework that can accommodate different distance metrics and behave robustly when outliers are present in the dissimilarities. 

We restrict the dissimilarity measure $d_{i,j}$ to be strictly positive and assume that it follows a truncated distribution:
\begin{equation}\label{dij_general}
d_{i,j} \sim g\left(\delta_{i,j}\right)I\left(d_{i,j}>0\right),\qquad   i\neq j,\; i,j = 1,\ldots,n,
\end{equation}
where $I(\cdot)$ is an indicator function. The true dissimilarity measure $\delta_{i,j}$ is modeled as the distance between object $i$ and $j$ using the dissimilarity metric $\mathcal{D}$:
\begin{equation}\label{distance metric}
\delta_{i,j} = \mathcal{D}(\mathbf{x}_i, \mathbf{x}_j).
\end{equation}
The GBMDS framework we propose is general in nature. The various GBMDS models differ from one another based on the choice of dissimilarity metric $\mathcal{D}$ and distribution function $g$.

Compared with the BMDS framework proposed in \citet{oh2001bayesian}, we do not restrict $d_{i,j}$ to be accompanied by Gaussian errors. The previous BMDS framework may be inadequate when dealing with dissimilarity measures that are subject to random errors or those that are non-Euclidean in nature. In addition, the assumption of utilizing a truncated Gaussian distribution to model the errors is inadequate in the presence of outliers.
The presence of outliers can lead to increased uncertainty surrounding unobserved dissimilarities ($\delta_{i,j}$'s) beyond what can be accounted for by the tails of Gaussian distributions. We will refer to the framework in \citet{oh2001bayesian} as the standard BMDS throughout this paper. 

\subsubsection{Dissimilarity metrics}
The standard choice of dissimilarity metric $\mathcal{D}$ on $\mathbb{R}^{p}$ is the Euclidean metric ($L^2$ norm): $ \mathcal{D}(\mathbf{x}_i, \mathbf{x}_j) = \| \mathbf{x}_i - \mathbf{x}_j \|_{2} = \sqrt{\sum_{k = 1}^{p} (x_{i,k} - x_{j,k})^2}$.  It is often used in MDS when the dissimilarity matrix satisfies the metric axioms and has a well-defined Euclidean interpretation.

We generalize the standard BMDS by considering cases where the dissimilarity matrix may not have a well-defined Euclidean interpretation. In this case, we can consider candidate models with non-Euclidean dissimilarity metrics. For example, Cosine metric is defined as $\mathcal{D}(\mathbf{x}_i, \mathbf{x}_j) = 1 - \left( \sum_{k = 1}^{p} x_{i,k} x_{j,k} \right) / \left(\sqrt{\sum_{k = 1}^{p} x_{i,k}^2} \sqrt{\sum_{k = 1}^{p} x_{j,k}^2} \right)$. This study utilizes the Cosine metric for text analysis using non-negative word frequencies, therefore, the Cosine metric of interest has a range from 0 to 1. 

In our GBMDS framework, a variety of distributions can be considered for $g$, including both symmetric and skewed distributions. Symmetric distributions, such as Gaussian or Student's $t$-distributions, are suitable in some cases, while in other scenarios, a skewed distribution is more appropriate. In what follows, we will focus on the truncated skewed Gaussian distribution.  This distribution is a suitable choice when the errors are skewed.  Then, we will proceed to investigate the truncated Student's $t$-distribution.  This distribution is deemed a suitable choice for robust estimation when outliers exist.

\subsubsection{GBMDS with truncated skewed Gaussian distribution}

We consider the possibility that some dissimilarities are both skewed and positive. We denote the model with truncated skewed Gaussian distribution as $\mathcal{M}_{TSN}$ and model the dissimilarity $d_{i,j}$ as follows:
\begin{center}
$d_{i,j} | \mathcal{M}_{TSN} \sim  \mathcal{TSN}\left(\delta_{i,j}, \sigma^{2}, \psi \right)I\left(0<d_{i,j}<U\right),\qquad   i\neq j,\; i,j = 1,\ldots,n$,
\end{center}
where $\sigma^{2} \in \mathbb{R}^+$ is the squared scale parameter, $\psi \in \mathbb{R}$ is the shape parameter, and $U$ is the upper bound. The truncated Gaussian distribution is recovered when $\psi$ is zero. As the absolute value of $\psi$ grows, the absolute skewness of the distribution also increases, with negative $\psi$ producing a left-skewed distribution and positive $\psi$ generating a right-skewed distribution. Generally, the upper bound $U$ for the truncated skewed Gaussian distribution can be set to infinity. However, for certain non-Euclidean dissimilarity metrics, it is necessary to impose an upper bound on $d_{i,j}$ to constrain the dissimilarities within a specific range. For example, when using the Cosine metric for text data, which ranges from 0 to 1, a natural choice for the upper bound is $U=1$. 

 Let the latent variables $\textbf{X} = \left\{ \mathbf{x}_1, \ldots, \mathbf{x}_n\right\}$. Under $\mathcal{M}_{TSN}$, for a given matrix of dissimilarities $\textbf{D}$, the likelihood function $L$ with respect to $\textbf{X}$, $\sigma^2$, and $\psi$, can be written as:
\begin{align*} \label{likelihood_M_TSN} 
   L\left(\textbf{X}, \sigma^{2}, \psi; \textbf{D}, \mathcal{M}_{TSN}\right) 
      & \propto \left(\sigma^{2}\right)^{-\frac{m}{2}} \prod_{\substack{i, j = 1 \\ i > j}}^n \left\{ F_{\delta_{i,j}, \sigma, \psi}\left(U\right) - F_{\delta_{i,j}, \sigma, \psi}\left(0\right) \right\}^{-1} \\ &
      \times \exp\left\{-\frac{1}{2\sigma^{2}} SSR\right\} \times \prod_{\substack{i, j = 1 \\ i > j}}^n \Phi \left( \psi \frac{d_{i,j} - \delta_{i,j}}{\sigma} \right), 
      \numberthis
\end{align*}
where $F_{\delta_{i,j}, \sigma, \psi}(\cdot)$ is the cdf of skewed Gaussian distribution with location $\delta_{i,j}$, scale $\sigma$, and shape $\psi$, $SSR = \sum_{i > j}\left(d_{i,j}-\delta_{i,j}\right)^{2}$ is the sum of squared residuals, $\Phi(\cdot)$ is the standard Gaussian cdf, and $m=n(n-1)/2$ is the total number of dissimilarities for $n$ objects.

\subsubsection{GBMDS with truncated Student's $t$-distribution}

To further relax the assumption of constant Gaussian error variance in the dissimilarity, we introduce the model with truncated Student's $t$-distribution. Consequently, we can accommodate different degrees of uncertainty associated with dissimilarity by using different error variances. The $t$-distribution is often used as an alternative to the Gaussian distribution as a more robust model to fit data with heavier tails \citep{lange1989robust, lin2019bayesian}. In many applications, the outliers add more uncertainty around the tails of the dissimilarity measures. Fitting the truncated $t$-distribution provides a longer tail. 

The $t$-distribution can be written in the form of its scale mixtures of Gaussian representation to demonstrate its robustness property:
\begin{equation}\label{t_robust}
    t_{\nu}\left(x; \mu, \sigma^{2}\right) = \int_0^{\infty} \mathcal{N}\left(x; \mu, \frac{\sigma^{2}}{\zeta}\right) Gamma\left(\zeta; \frac{\nu}{2}, \frac{\nu}{2}\right) d\zeta.
\end{equation}
Equation (\ref{t_robust}) indicates that if a random variable $x$ follows a $t$-distribution with mean $\mu$, variance $\sigma^2$, and degrees of freedom $\nu$, then conditioning on $\zeta \sim \textit{Gamma}\left(\nu/2, \nu/2\right)$, $x$ follows a Gaussian distribution with parameters $\mu$ and $\sigma^2 / \zeta$. The $t$-distribution downweighs the observations which are disparate from the majority under the Gaussian distribution. This means that observations that are outliers or significantly different from the majority of the data will have less influence on the overall distribution in the $t$-distribution compared to the Gaussian distribution.

We denote the model with truncated $t$-distribution as $\mathcal{M}_{TT}$. $\mathcal{M}_{TT}$ models the dissimilarity $d_{i,j}$ as follows: 
\begin{align*}
\zeta_{i,j} &\sim \textit{Gamma}\left(\nu/2, \nu/2\right), \\
d_{i,j} | \mathcal{M}_{TT} &\sim \mathcal{N}\left(\delta_{i,j}, \sigma^{2}/\zeta_{i,j}\right)I\left(0<d_{i,j}<U\right), \qquad   i\neq j,\; i,j = 1,\ldots,n.
\end{align*}

Under $\mathcal{M}_{TT}$, for a given matrix of dissimilarities $\textbf{D}$, the likelihood function $L$ with respect to $\textbf{X}$, $\sigma^2$, and $\zeta_{i,j}$, can be written as:
\begin{align*} \label{likelihood_M_T} 
    L & \left(\textbf{X}, \sigma^{2}, \zeta_{i,j}; \textbf{D}, \mathcal{M}_{TT}\right) 
      \propto \left(\sigma^{2}\right)^{-\frac{m}{2}}   \exp\left\{\frac{1}{2}\sum_{\substack{i, j = 1 \\ i > j}}^n\log\left(\zeta_{i,j}\right)-\frac{1}{2\sigma^{2}}\sum_{\substack{i, j = 1 \\ i > j}}^n\zeta_{i,j}\left(d_{i,j}-\delta_{i,j}\right)^{2} \right. \\ & \left.
      - \sum_{\substack{i, j = 1 \\ i > j}}^n\log \left( \Phi\left(\frac{\left(U -\delta_{i,j}\right) \sqrt{\zeta_{i,j}}}{\sigma}\right)
      - \Phi\left(-\frac{\delta_{i,j} \sqrt{\zeta_{i,j}}}{\sigma}\right)\right)\right\},
      \numberthis
\end{align*}
where $\Phi(\cdot)$ and $m=n(n-1)/2$ are defined as in the model $\mathcal{M}_{TSN}$.

\subsection{Bayesian inference}

\subsubsection{Prior distributions} \label{sec:prior}

Under the Bayesian framework, the prior distributions for the unknown parameters need to be specified in advance. We first introduce the prior distributions for model $\mathcal{M}_{TSN}$ with unknown parameters $\textbf{x}_i$, $\sigma^{2}$, and $\psi$. We assume prior independence among parameters. For the prior of $\textbf{x}_i$, we choose a multivariate Gaussian distribution with mean $\textbf{0}$ and a diagonal covariance matrix $\Lambda = \text{diag}(\lambda_{1},\ldots, \lambda_{p})$. In other words, $\textbf{x}_i \sim \mathcal{N}\left(\textbf{0}, \Lambda\right)$, independently for $i = 1,\ldots,n$. For the elements along the diagonal covariance matrix, we assume an inverse Gamma distribution for the hyperprior distribution, i.e., $\lambda_{k} \sim \mathcal{IG}\left(\alpha, \beta_{k}\right)$, independently for $k = 1,\ldots,p$. Under this assumption, we posit independence among the latent variables, resulting in a simpler model with reduced hyperparameters and improved interpretability. For $\sigma^2$, we use an inverse Gamma distribution for the prior distribution, i.e., $\sigma^2 \sim \mathcal{IG}\left(a, b\right)$. For $\psi$, we choose a diffuse prior, i.e., $\psi \sim \mathcal{U}\left(c, d\right)$. We denote the prior distributions of the unknown parameters as $\pi\left(\textbf{X} | \Lambda \right)$, $\pi\left(\sigma^{2}\right)$, $\pi\left(\psi\right)$, and $\pi\left(\Lambda\right)$.

Next, we specify the prior distributions for model $\mathcal{M}_{TT}$ with unknown parameters $\textbf{x}_i$, $\sigma^{2}$, and $\zeta_{i,j}$. We use the same settings for the prior distributions of $\textbf{x}_i$ and $\sigma^{2}$ as in model $\mathcal{M}_{TSN}$. In addition, we use $Gamma\left(\nu/2, \nu/2\right)$ as the prior distribution for $\zeta_{i,j}$.

\subsubsection{Posterior distributions} \label{sec:posterior}
For simplicity, we introduce a new notation $\particle$ to represent all the latent variables $\hidden$ and the unknown parameters $\parameter$.  Note that the parameters $\parameter$ vary across different models. In the model with truncated skewed Gaussian distribution, $\parameter$ includes $\sigma^{2}$, $\psi$, and $\Lambda$. In the model with truncated Student's $t$-distribution, $\parameter$ includes  $\sigma^{2}$, $\zeta$, and $\Lambda$.

In the Bayesian framework, our interest lies in the posterior distribution of $\particle$ given the dissimilarity matrix $\textbf{D}$, denoted as:
\begin{equation}\label{normalized_posterior}
\pi\left(\particle | \textbf{D}\right) = \frac{\gamma\left(\particle | \textbf{D}\right)}{Z} \propto L(\particle; \textbf{D}) \, \pi(\particle),
\end{equation}
where $\gamma(\particle | \textbf{D}) = L(\particle; \textbf{D})\pi(\particle)$ denotes the unnormalized posterior distribution, $L(\particle; \textbf{D})$ is the likelihood function, $\pi(\particle)$ is the prior on the parameters, and $Z = p(\textbf{D}) = \int{\gamma(\particle | \textbf{D}) \, d\particle}$ is the marginal likelihood. The likelihood functions are specified in Equations \ref{likelihood_M_TSN} and \ref{likelihood_M_T} for $\mathcal{M}_{TSN}$ and $\mathcal{M}_{TT}$, respectively. The prior distributions are described in the previous subsection. Since the normalizing constant $Z$ is intractable, we will use Monte Carlo methods to approximate the posterior distributions, which will be detailed in Section \ref{sec:GBMDS_methodology}.

\subsubsection{Adaptive Bayesian inference} \label{sec:adaptive_model}

We propose an adaptive Bayesian inference framework for two key scenarios.  First, in online inference, we sequentially update posterior estimates as new data arrive to allow for real-time updates. This approach is particularly useful when results need to be refined incrementally rather than recomputed from scratch. The main idea is to use the posterior distribution from a previous iteration to initialize the next. Second, for large datasets, instead of processing all observations at once, we can partition the data into smaller batches and perform inference sequentially. In BMDS, such an approach is especially beneficial as it enables dynamic visualization updates while managing computational efficiency. 
Here, we provide an overview of the adaptive Bayesian inference process. A detailed explanation of the adaptive mechanism is presented in Section \ref{sec:ASMC_adaptive}.

Let $\hidden^{(0)}$ be the hidden variables associated with $\textbf{Z}^{(0)} = \{\mathbf{z}_1, \ldots, \mathbf{z}_{n_0}\}$, and  $\hidden^{(1)}$ be the hidden variables associated with $\textbf{Z}^{(1)} = \{\mathbf{z}_{n_0+1}, \ldots, \mathbf{z}_{n_0+n_1}\}$. Here, $n_0$ refers to the dimension of the original observations, and $n_1$ refers to the incremental dimension of the additional observations. Given dissimilarity metric $\mathcal{D}$, dissimilarities $\data^{(0)}$ is obtained from $\textbf{Z}^{(0)}$ and $\data$ is obtained from $\textbf{Z} = \textbf{Z}^{(0)}\cup \textbf{Z}^{(1)}$.

In this case, $\particle$ is composed of three parts, $\hidden^{(0)}$, $\hidden^{(1)}$, and $\parameter$.  The posterior distribution of $\particle$ can be rewritten as: 
\begin{equation}\label{normalized_posterior_incr}
\pi\left(\hidden^{(0)}, \hidden^{(1)}, \parameter | \data \right)  \propto L\left(\hidden^{(0)}, \hidden^{(1)}, \parameter; \textbf{D}\right ) \pi\left(\hidden^{(0)} | \Lambda ^{(0)}\right) \pi\left(\hidden^{(1)} | \Lambda ^{(1)}\right) \pi\left(\parameter\right). 
\end{equation}
The adaptive Bayesian inference concerns the inference of $\pi(\hidden^{(0)}, \hidden^{(1)}, \parameter | \data )$ using the previous inference for $\pi(\hidden^{(0)}, \parameter | \data^{(0)} )$ when dissimilarity data increase from $\data^{(0)}$ to $\data$.

When the previous dissimilarity matrix $\data^{(0)}$ is not available, we denote $\data^{(0)}=\emptyset$ and $\hidden^{(0)}=\emptyset$. With our notation, the posterior distribution in Equation \ref{normalized_posterior} is a special case of 
Equation \ref{normalized_posterior_incr} when $\data^{(0)}=\emptyset$ and $\hidden^{(0)}=\emptyset$. Therefore, we will only focus on the 
adaptive Bayesian inference with Monte Carlo methods for Equation \ref{normalized_posterior_incr} in Section \ref{sec:GBMDS_methodology}.

\subsection{Model comparison} \label{sec:bayes_factor}

As described in the previous section, the function $g$ can take different forms. In most cases, the optimal form of $g$, the number of significant attributes $p$, and the suitable dissimilarity metrics are unknown. In this section, we approach the problem of comparing a discrete set of Bayesian models with the Bayes factor. Consider two models $\mathcal{M}_1$ and $\mathcal{M}_2$ with different likelihoods and corresponding sets of parameters $\particle_1$ and $\particle_2$. In the context of this paper, $\mathcal{M}_1$ and $\mathcal{M}_2$ would correspond to two competing models. Examples of competing models could be $\mathcal{M}_{TSN}$ versus $\mathcal{M}_{TT}$, or one model under a different choice of dimension $p$. The Bayes factor is defined as the ratio of the posterior odds to the prior odds. When the two models have equal prior probability, the Bayes factor reduces to the ratio of two marginal likelihoods, given by:
\begin{equation*}
    \text{BF}_{12} =
    \frac{\int \pi(\particle_1 | \mathcal{M}_1) L(\particle_1; \textbf{D}, \mathcal{M}_1) \,d\particle_1}{\int \pi(\particle_2 | \mathcal{M}_2) L(\particle_2; \textbf{D},  \mathcal{M}_2) \,d\particle_2} = 
    \frac{p\left(\textbf{D} | \mathcal{M}_1\right)}{p\left(\textbf{D} | \mathcal{M}_2\right)}.
\end{equation*}
The Bayes factor quantifies the support for one model over another; a value greater than 1 supports Model 1, while a value less than 1 supports Model 2. Following the guidelines in \citet{kass1995bayes}, a Bayes factor between 1 and 3 provides negligible evidence, while a value between 3 and 20 suggests positive evidence. A Bayes factor between 20 and 150 indicates strong evidence, and a value exceeding 150 is considered very strong evidence.

A typical challenge for using the Bayes factor is the computation of the marginal likelihood estimates, especially for MCMC-based methods. Marginal likelihood estimation is not straightforward in standard MCMC, often necessitating additional, complex sampling procedures or post-processing \citep{chib2001marginal, skilling2004nested, robert2009computational}. Furthermore, many MCMC-based estimators, such as the harmonic mean, can suffer from infinite variance or significant finite-sample bias. In contrast, standard SMC algorithms produce unbiased marginal likelihood estimates naturally as a byproduct of the sequential sampling process \citep{del2004feynman}. While adaptive SMC strategies may introduce negligible bias, they generally avoid the complex implementation hurdles associated with MCMC-based evidence estimation  \citep{zhou2016toward, wang2020annealed}. 

On the other side, in the frequentist view, STRESS is a commonly used measure of fit for the object configuration problem \citep{kruskal1964multidimensional}. STRESS value is defined as
\begin{equation*}
    \text{STRESS} = \sqrt{\frac{\sum_{i > j} \left(d_{i,j} - \hat{\delta}_{i,j}\right)^2}{\sum_{i > j} d^2_{i,j}}}, \qquad  i,j = 1,\ldots,n,
\end{equation*}
where $\hat{\delta}_{i,j}$ is the distance found from the estimated object configuration. 
A smaller STRESS value indicates a better fit. 

In this work, we select the optimal Bayesian model and dimension $p$ using marginal likelihood estimates. Furthermore, we compare the performance of CMDS and GBMDS using the STRESS value.

\subsection{Identifiability in multidimensional scaling} \label{sec:identifiability}
Similar to other dimensional reduction methods, identification issues arise in the posterior inference of GBMDS. For instance, the center and direction of the estimated points can be arbitrary. To address this, we standardize all posterior samples of $\textbf{x}_i$'s using Procrustes transformations \citep{goodall1991procrustes}. This method aligns configurations using a least-squares criterion by a combination of scaling, rotation, reflection, and translation. After the Procrustes transformation, we construct credible regions from the adjusted posterior samples of $\textbf{x}_i$ to quantify uncertainty.

\section{Adaptive Bayesian Inference using Annealed SMC}
\label{sec:GBMDS_methodology}

\subsection{Intermediate distributions and particle initialization}
\label{sec:ASMC_initialization}

To conduct the Bayesian inference for the posterior distribution in Equation \ref{normalized_posterior_incr}, we propose to design an artificial sequence of annealing intermediate target distributions following the ideas from the SMC literature \citep{neal2001annealed, del2006sequential, Moral2007, wang2020annealed}.  Specifically, we create a sequence of annealing intermediate target distributions $\{\pi_r(\particle)\}_{0 \leq r\leq R}$, such that
\begin{equation}\label{seq_distribution}
\pi_r\left(\particle\right) \propto \gamma_r\left(\particle\right) = \left(L\left(\particle; \textbf{D} \right)\pi\left(\particle\right)\right)^{\tau_r} \times \tilde\pi\left(\particle\right)^{1-\tau_{r}} ,
\end{equation}
where $\tilde\pi(\particle)$ is a \emph{reference distribution} that is generally easy to sample from \citep{fan2011choosing}, and $0 = \tau_0 < \tau_1 < \cdots < \tau_R = 1$ is a sequence of annealing parameters. If $\tau_r$ is zero, the distribution becomes the reference distribution $\tilde\pi(\particle)$. At the other extreme, the distribution is the posterior distribution of interest when the power $\tau_r$ equals 1. 

In our model, $\particle$ is a vector of all the variables in $\hidden^{(0)}$, $\hidden^{(1)}$, and $\parameter$. 
The reference distribution can be specified for $\hidden^{(0)}$, $\hidden^{(1)}$ and $\parameter$ independently: 
\begin{equation} \label{reference_distribution}
\tilde\pi\left(\particle \right) = \tilde\pi_0\left(\textbf{X}^{(0)}\right) 
\tilde\pi_1\left(\textbf{X}^{(1)}\right) \tilde\pi_{\theta}\left(\parameter\right). 
\end{equation}
Preferably, the reference distributions should possess properties that allow for convenient sampling and proximity to the modes of the target distribution. For simplicity, we choose the reference distribution for $\parameter$ to be its prior distribution, i.e. $\tilde\pi_{\theta}(\parameter) = \pi(\parameter)$, and the reference distributions for $\hidden^{(1)}$ to be its prior, $\tilde\pi_1\left(\textbf{X}^{(1)}\right) = \pi\left(\textbf{X}^{(1)} | \Lambda ^{(1)} \right)$; the reference distributions for $\hidden^{(0)}$, denoted $\tilde\pi_0\left(\textbf{X}^{(0)}\right)$,  is set to be a Gaussian distribution, detailed in Section \ref{sec:ASMC_adaptive}.

With a small value of $\tau_r$, the intermediate target distribution is closer to the reference distribution. For parameters that rely on the prior distribution as the reference distribution, smaller $\tau_r$ can result in flatter intermediate target distributions that facilitate the movement between various modes. The samples are coerced into the posterior distribution as we slowly increase the annealing parameter $\tau_r$. The initialization of particles is summarized in Algorithm \ref{alg:anneal_SMC_initialization}.

\begin{algorithm}[htbp]
\caption{\texttt{Particle\_Initialization}}
\label{alg:anneal_SMC_initialization}
\linespread{1.05}\selectfont
\small
\DontPrintSemicolon
\SetKwInOut{Input}{Input}\SetKwInOut{Output}{Output}
\Input{(a) Dissimilarity $\textbf{D}^{(0)}, \textbf{D}$; 
(b) Reference distributions $\tilde\pi_0, \tilde\pi_1,$ and $\tilde\pi_{\theta}$ for $\hidden^{(0)}, \hidden^{(1)}$, and $\parameter = \{\sigma^{2}, \Lambda, \psi, \zeta\}$, respectively. (c) Number of particles $K$.}
\Output{Initializations of $K$ particles: $\{\particle_{0, k}\}_{k=1}^K$.}
\BlankLine
\For{$k \in \{1, 2, \ldots, K\}$}{
\If{$\textbf{D}^{(0)} \neq \emptyset$}{ \label{alg:adaptive_start}
Initialize particles of $\textbf{X}^{(0)}$: $\textbf{X}^{(0)}_{0, k} \sim \tilde\pi_0(\cdot)$. 
}
Initialize particles of $\textbf{X}^{(1)}$: $\textbf{X}^{(1)}_{0, k} \sim \tilde\pi_1(\cdot)$. 
\BlankLine
Initialize particles of parameters: 
$\{{\sigma^{2}_{0, k}}, \Lambda_{0, k}, \psi_{0, k}, \zeta_{0, k}\}\sim \tilde\pi_{\theta}(\cdot)$. \label{alg:adaptive_end}
\BlankLine
Set $\particle_{0, k} \leftarrow \left\{ \textbf{X}^{(0)}_{0, k}, \textbf{X}^{(1)}_{0, k}, \sigma^{2}_{0, k}, \Lambda_{0, k}, \psi_{0, k}, \zeta_{0, k} \right\}$.
}
\end{algorithm}

\subsection{Annealed SMC}
\label{sec:ASMC_main}
This section will consider the annealed SMC algorithm for a fixed dimension where the previous dissimilarity matrix $\data^{(0)}$ is unavailable. We will introduce in Algorithm \ref{alg:anneal_SMC_main} the annealed SMC algorithm along with the adaptive mechanism for choosing the annealing sequence. The annealed SMC algorithm approximates the posterior distribution $\pi(\particle | \textbf{D})$ in $R$ steps. At each step $r$, we approximate $\pi_r(\cdot)$ using a total of $K$ particles. Each particle $\particle_{r, k}$ is associated with a positive weight. Let $w_{r, k}$ denote the unnormalized weight for particle $\particle_{r, k}$ and let $W_{r, k}$ denote the corresponding normalized weight. The normalization is performed by $W_{r, k} = w_{r, k}/\sum^K_{k=1} w_{r, k}$.

We start by sampling initial particles from the reference distributions. Then, the annealed SMC algorithm iterates between \emph{reweighting, propagating}, and \emph{resampling}. The details of the three steps in the annealed SMC algorithm are given as follows.

\textit{Step 1. Weight Update} 

The incremental importance weight for particle $k$ at iteration $r$ is
\begin{equation}\label{incr_weight}
\tilde{w}_{r, k} = \frac{\gamma_{r}\left(\particle_{r, k}\right) \times \kappa^{-}\left(\particle_{r, k} , \particle_{r-1, k}\right)}{\gamma_{r-1}\left(\particle_{r-1, k}\right) \times \kappa^{+}\left(\particle_{r-1, k} , \particle_{r, k}\right)},
\end{equation}
where the forward kernel  $\kappa^{+}(\particle_{r-1, k}, \particle_{r, k})$ is a $\pi_r$-invariant Metropolis-Hastings kernel, and $\kappa^{-}(\particle_{r, k}, \particle_{r-1, k})$ is the backward kernel \citep{del2006sequential}. The selection of the backward kernel is crucial as it will affect the variance of the normalized weights. A convenient backward kernel that allows easy computation of the weight is
\begin{equation}\label{back_kernel}
\kappa^{-}\left(\particle_{r, k} , \particle_{r-1, k}\right) = \frac{\gamma_{r}\left(\particle_{r-1, k}\right) \times \kappa^{+}\left(\particle_{r-1, k} , \particle_{r, k}\right)}{\gamma_{r}\left(\particle_{r, k}\right)}.
\end{equation}

This approach simplifies the evaluation of weights since we do not need point-wise evaluations of the backward and forward kernels. The incremental importance weight becomes 
\begin{equation}\label{incr_weight_simplified}
\tilde{w}_{r, k} = \left[\frac{L\left(\particle_{r-1, k}; \textbf{D} \right)\pi(\particle_{r-1,k})}{\tilde{\pi}_0(\particle_{r-1,k})} \right]^{\tau_{r} - \tau_{r-1}}.
\end{equation}
The weight update function for particles at iteration $r$ is 
\begin{align*}\label{weight}
    W_{r, k} \propto w_{r, k} = w_{r-1, k}  \tilde{w}_{r, k}. 
\end{align*}

Note the weight update function only depends on the particles at the previous iteration. This is implemented in Line \ref{alg:weight_update} of Algorithm \ref{alg:anneal_SMC_main}. 

\textit{Step 2. Particle Propagation} 

We sample the new particles $\particle_{r, k}$ from $\pi_r$-invariant Metropolis-Hastings kernels. At each step $r$, we have the option to sample either all parameters or a subset of the parameters, depending on the desired trade-off between computational efficiency and effective exploration of the parameter space. In particular, for a high-dimensional parameter $\left\{ \mathbf{x}_1, \ldots, \mathbf{x}_n\right\}$, updating a subset of it can help mitigate the risk of low acceptance rates. This approach allows for more targeted exploration of the parameter space and potentially results in faster mixing. 
The full conditional distributions for  latent variables $\left\{ \mathbf{x}_1, \ldots, \mathbf{x}_n\right\}$, and parameters $\lambda_k$, $\sigma^2$, and $\zeta_{i,j}$ are presented below. In each full conditional distribution, we use $| \cdot$ to denote conditioning on the data and all other parameters. A detailed description of sampling methods is given in \textit{Supplementary}.

The full conditional distribution for $\lambda_k$ is
\begin{equation}\label{eqn:lambda_full_conditional_distribution}
    \lambda_k | \cdot \sim IG(\alpha + n/2, \beta_k + \tau_{r}v_k/2),
\end{equation}
where $v_k$ is the quantity for which $v_k/n$ is the sample variance of the $k$th coordinates of $\textbf{x}_i$'s.

The full conditional distributions of $\left\{ \mathbf{x}_1, \ldots, \mathbf{x}_n\right\}$, $\sigma^2$ and $\psi$ do not admit closed forms, a random walk Metropolis-Hastings step is implemented with the Gaussian proposal densities.

For $\mathcal{M}_{TSN}$,
\begin{align}
    \gamma_r \left(\left\{ \mathbf{x}_1, \ldots, \mathbf{x}_n\right\} | \cdot\right) & \propto \exp \left\{ -\tau_r \left( A + \frac{1}{2} \sum_{i = 1}^{n} \textbf{x}_i^{\top} \Lambda^{-1} \textbf{x}_i \right) \right\}, \\
    \gamma_r \left(\sigma^2 | \cdot\right) & \propto \sigma^{-2(a+1)} \exp \left\{ -\tau_r \left( A + \frac{b}{\sigma^2} \right) \right\},
\end{align}
where 

$A = \frac{1}{2\sigma^{2}} \: SSR +\frac{m}{2} \log \left(  \sigma^{2} \left(F_{\delta_{i,j}, \sigma, \psi}(U) - F_{\delta_{i,j}, \sigma, \psi}(0)\right) \right)  - \sum\limits_{i > j} \log \left( \Phi \left( \psi \frac{d_{i,j} - \delta_{i,j}}{\sigma} \right) \right)$.

For $\mathcal{M}_{TT}$, 
\begin{align}
    \gamma_r \left( \left\{ \mathbf{x}_1, \ldots, \mathbf{x}_n\right\} | \cdot\right) & \propto \exp\left\{-\tau_r \left( C  + \frac{1}{2} \sum_{i = 1}^{n} \textbf{x}_i^{\top} \Lambda^{-1} \textbf{x}_i \right) \right\}, \\
    \gamma_r \left( \sigma^2 | \cdot \right) & \propto \sigma^{-m} \exp\left\{-\tau_r \left( C + \frac{b}{\sigma^2} \right) \right\},
\end{align}
where $C = \frac{1}{2\sigma^{2}}\sum_{i > j}\zeta_{i,j}\left(d_{i,j}-\delta_{i,j}\right)^{2}+ \sum\limits_{i > j}\log \left( \Phi\left(\frac{\left(U -\delta_{i,j}\right) \sqrt{\zeta_{i,j}}}{\sigma}\right)
      - \Phi\left(-\frac{\delta_{i,j} \sqrt{\zeta_{i,j}}}{\sigma}\right)\right)$.

For model $\mathcal{M}_{TT}$, the full conditional distribution for $\zeta_{i,j}$ is
\begin{equation}\label{eqn:zeta_full_conditional_distribution}
    \zeta_{i,j} | \cdot \sim Gamma((\tau_{r}+\nu)/2, \tau_{r}(d_{i,j}-\delta_{i,j})^{2}/(2\sigma^2) + \nu/4).
\end{equation}

\textit{Step 3. Particle Resampling} 

To alleviate the issue of particle degeneracy, where all normalized weights approach to 0 except for one, we resample the particles to replicate those with high weights and discard those with low weights.  Popular resampling schemes include, but are not limited to, multinomial, systematic, stratified, and residual resampling \citep{douc2005comparison}. We employ multinomial resampling. Although resampling alleviates particle degeneracy, performing it at every iteration introduces excess Monte Carlo variance into the estimator. 
Therefore, we perform resampling only when the degeneracy of the particles exceeds a specific threshold $\epsilon$. We monitor this using the effective sampling size (ESS) \citep{kong1992note}:
\begin{equation}\label{ESS}
\text{ESS} = \frac{1}{\sum^{K}_{k = 1} \left(W_{r, k}\right)^2}.
\end{equation}
The relative effective sample size (rESS), defined as  $\text{rESS} = \text{ESS} /K$,  normalizes this metric to the interval [0,1]. 

\begin{algorithm}[htbp]
\caption{\texttt{Annealed\_SMC}}
\label{alg:anneal_SMC_main}
\linespread{1.05}\selectfont
\small
\DontPrintSemicolon

\SetKwInOut{Input}{Input}\SetKwInOut{Output}{Output}
\Input{(a) Initialization of $K$ particles: $\{\particle_{0, k}\}_{k=1}^K$; (b) Priors $\pi(\particle)$ and reference distributions $\tilde\pi(\particle)$ for $\particle = \{\left\{ \mathbf{x}_1, \ldots, \mathbf{x}_n\right\}, \sigma^{2}, \Lambda, \psi, \zeta\}$; (c) Likelihood function $L(\particle; \textbf{D})$; (d) rCESS threshold $\phi$; (e) Resampling threshold $\epsilon$.}
\Output{(a) Particle population: $\{(\particle_{R, k}, W_{R, k})\}_{k=1}^K$; (b) Marginal likelihood estimates: $\widehat{Z}_R$; (c) Total SMC iterations: $R$; (d) Sequence of annealing parameter: $\{\tau_r\}_{r=0}^R$.}
\BlankLine

Initialize SMC iteration index: $r \leftarrow 0$, initialize annealing parameter: $\tau_0 \leftarrow 0$, initialize marginal likelihood estimate: $\widehat{Z}_0 \leftarrow 1$, load initial particles $\{\particle_{0, k}\}_{k=1}^K$.

\For{$k \in \{1, 2, \ldots, K\}$}{$w_{r,k} = 1$, $W_{r,k} = 1/K$.}

\For{$r \in \{1, 2, \ldots\}$}{
 \For{$k \in \{1, 2, \ldots, K\}$}{
    Express the incremental importance weights as a function of $\tau_r$:
    $\tilde{w}_{r, k} = \left[\frac{L\left(\particle_{r-1, k}; \textbf{D}\right)\pi(\particle_{r-1,k})}{\tilde{\pi}(\particle_{r-1,k})} \right]^{\tau_{r} - \tau_{r-1}}.$

    }
    Determine the next annealing parameter $\tau_r$ using bisection method with:
    $f(\tau) = \text{rCESS}_r(W_{r-1, \cdot}, \tilde w_{r,\cdot}) = \phi$. \label{alg:anneal}
   
    \For{$k \in \{1, 2, \ldots, K\}$}{
        Compute unnormalized weights: $w_{r, k} = w_{r-1, k} \, \tilde{w}_{r,k}$. \label{alg:weight_update}
        
        Normalize weights: $W_{r, k} = w_{r, k}/(\sum^K_{k=1} w_{r, k})$.
        
        Sample particles $\particle_{r, k}$ from $\pi_r$-invariant Metropolis-Hastings kernels using (\ref{eqn:lambda_full_conditional_distribution}) -(\ref{eqn:zeta_full_conditional_distribution}).
    }
    
    Update marginal likelihood estimates $\widehat{Z}_r = \widehat{Z}_{r-1}\, \sum^K_{k=1} W_{r-1,k}{\tilde{w}}_{r, k}$.
    
    \eIf{$\tau_r \geq 1$}{
        Set  $\tau_r \leftarrow  1$. Repeat Steps 5–6 and 8–12. 
        
        The total number of SMC iterations $R\leftarrow r$.         
        
        Return $R$, $\{\tau_r\}_{r=0}^R$, $\{(\particle_{R, k}, W_{R, k})\}_{k=1}^K$, and $\widehat{Z}_R$.
        }{
        \If{particle degeneracy is too severe, i.e. rESS $< \epsilon$}{
            Resample the particles, denoted $\{\particle_{r, k}\}_{k=1}^K$; \BlankLine
            Reset particle weights:  $w_{r,k} = 1$, $W_{r,k} = 1/K$.
            }
        }
    }
\end{algorithm}

The annealed SMC algorithm produces a set of particles. After the extra resampling step in the end, the output of the annealed SMC algorithm contains a list of $K$ particles with equal weight. These particles can be used for the posterior approximation and for constructing the visualization in the lower-dimensional space. To find the Bayesian estimate of $\textbf{X}$, we take an approximate posterior mode of $\left\{ \mathbf{x}_1, \ldots, \mathbf{x}_n\right\}$ as described in \citet{oh2001bayesian}. \citet{oh2001bayesian} observed that the term involving $SSR$ dominates the posterior density. Thus, the approximate posterior mode can be found by the values of $\left\{ \mathbf{x}_1, \ldots, \mathbf{x}_n\right\}$ that minimizes $SSR$ among all $K$ particles. The approximate posterior mode retrieves the relative positions of $\left\{ \mathbf{x}_1, \ldots, \mathbf{x}_n\right\}$, and they can be considered as the solution to the object configuration. Meaningful absolute positions of $\textbf{X}$ may be obtained from some suitable transformation defined by the users if needed.

Some challenges arise with MCMC-based approximations in the context of model comparison via marginal likelihood estimators, as discussed in \ref{sec:bayes_factor}. 
MCMC-based algorithms often require additional computational costs or complex post-processing to estimate the marginal likelihood separately. In contrast, model selection can be accomplished directly using the Bayes factor in the proposed annealed SMC algorithm, where the marginal likelihood is a natural byproduct of the sampling process. The estimated marginal likelihood is evaluated using the following formula:
$$\widehat{Z}_R = \prod_{r=1}^R \sum_{k=1}^K W_{r-1,k} \tilde{w}_{r,k}.$$
Moreover, these estimates are unbiased when using a fixed annealing sequence and nearly unbiased with an adaptive annealing
scheme. Previous studies have shown the advantages of SMC over MCMC for model comparison \citep{zhou2016toward, wang2020annealed}.

The sequence of intermediate target distributions, as defined in Equation \ref{seq_distribution}, is determined by the choice of the annealing sequence, $\{\tau_r\}$. Proper selection of the sequence of annealing parameters is one challenge in the annealed SMC. A large number of annealing parameters can improve the performance, but increase the computational cost. In order to ensure the proposed particles from the current iteration can effectively approximate the subsequent intermediate target distribution, it is necessary to transition smoothly from the reference distribution ($\tau_0 = 0$) to the posterior distribution ($\tau_R = 1$). 

We apply the adaptive annealing parameter scheme. The main idea is to select an annealing parameter $\tau$ such that we achieve a controlled increase in particle degeneracy. The particle degeneracy between two successive intermediate distributions is measured by the relative conditional effective sample size (rCESS) \citep{zhou2016toward},
\begin{equation}\label{rCESS}
\text{rCESS}_r\left(W_{r-1, \cdot}, \tilde{w}_{r, \cdot}\right) = \frac{\left(\sum^K_{k=1} W_{r-1, k} \tilde{w}_{r,k}\right)^2}{\sum^{K}_{k = 1} W_{r-1, k}\left(\tilde{w}_{r, k}\right)^2}.
\end{equation}
Values of rCESS range from $1/K$ to 1. With the $\tilde{w}_{r, k}$
in Equation \ref{incr_weight_simplified}, $\text{rCESS}_r$ is a decreasing function of $\tau_r$, where $\tau_r \in (\tau_{r-1}, 1]$. The value of rCESS over iterations is controlled by choosing the annealing parameter $\tau$ such that
\begin{equation}\label{phi}
f(\tau) = \text{rCESS}_r\left(W_{r-1, \cdot}, \tilde{w}_{r,\cdot}\right) = \phi,
\end{equation}
where $\phi \in (0, 1)$ is a tuning parameter that controls the length of the sequence $\tau_r$. Since there exists no closed-form solution for $\tau$ by solving $f(\tau) = \phi$, a bisection method is used to solve this one-dimensional search problem. The search interval is $\tau_r \in (\tau_{r-1}, 1]$. Given that $f$ is a continuous function with $f(\tau_{r-1}) - \phi > 0$ and $f(1) - \phi < 0$ (otherwise set $\tau_r = 1$), it follows that there must exist an intermediate point $\tau^*$ with $f(\tau^*) = \phi$. This is implemented in Line \ref{alg:anneal} of Algorithm \ref{alg:anneal_SMC_main}.

\subsection{Adaptive Bayesian inference}
\label{sec:ASMC_adaptive}

We have presented the annealed SMC algorithm for a fixed dataset. In this section, we describe a sequential mechanism to enable the algorithm to handle incremental batches of data. The complete algorithm is presented in Algorithm \ref{alg:anneal_SMC_adaptive}.

Suppose we have already obtained the posterior samples of $\hidden^{(0)}$ from $m_0$ old data points by running the annealed SMC algorithm, and an additional $m_1$ new data points become available. In this case, instead of running the annealed SMC algorithm from scratch using the combined dataset of size $n = m_0 + m_1$, we can utilize the information from the existing posterior samples to initialize  the particles for $\hidden^{(0)}$ and use the prior distribution to initialize samples for the new latent variables $\hidden^{(1)}$.

More specifically, inferences are updated sequentially as new batches of data arrive. When a new batch becomes available, the dissimilarity matrix $\data = \data^{(0)} \cup \data^{(1)}$ is calculated based on all available data including both old data of size $m_0$ and new data of size $m_1$.  The objective is to conduct inference for the joint posterior $\pi(\hidden^{(0)}, \hidden^{(1)}, \parameter | \data )$. To achieve this, we employ the strategy detailed in Section \ref{sec:ASMC_initialization} to initialize the particles of $\particle = \{\hidden^{(0)}, \hidden^{(1)}, \parameter\}$. 

When no previous data are available, we have $\data^{(0)}=\emptyset$, $\hidden^{(0)}=\emptyset$, $m_0=0$, and $n=m_1$. The reference distribution for the initial latent variables $\hidden^{(1)} = \left\{ \mathbf{x}_1, \ldots, \mathbf{x}_n\right\}$ is constructed using the solution from CMDS:
\begin{equation*}
    \textbf{x}_{i} \sim \mathcal{N}\left(\textbf{x}_{i}^{\text{CMDS}}, \rho \boldsymbol{I} \right), \text{ for } i = 1,\ldots,n,
\end{equation*}
where $\textbf{x}_{i}^{\text{CMDS}}$ is the result from fitting CMDS on $\data^{(1)}$. The variance $\rho$ is selected to concentrate particles around the CMDS solution. With a small value of the annealing parameter $\tau_r$, the intermediate target distributions remain close to this data-driven initialization.

When prior information exists ($\data^{(0)} \neq \emptyset$ and $\hidden^{(0)} \neq \emptyset$), we construct the reference distribution for the existing latent variables $\hidden^{(0)}$ using a Gaussian approximation derived from the previous posterior samples. Specifically: 
\begin{equation*}
    \textbf{x}_{i} \sim \mathcal{N}\left(\hat{\textbf{x}}_i, \hat{\boldsymbol{\Sigma}}  \right), \text{ for }  i = 1,\ldots,m_0,
\end{equation*}
where $\hat{\textbf{x}}_i$ is the posterior mode of $\textbf{x}_i$  and 
 $\hat{\boldsymbol{\Sigma}}$ represents the estimated covariance matrix computed from the previous particle set.  For the new incremental set $\hidden^{(1)}$ in $\particle$, we sample initial particles from the reference distribution, which is chosen to be its prior distribution for simplicity. This completes the specifications of the reference distributions in Algorithm \ref{alg:anneal_SMC_initialization}. Initialization of particles is implemented from Line \ref{alg:initialize_s} to Line \ref{alg:initialize_e} of Algorithm \ref{alg:anneal_SMC_adaptive}.

\begin{figure}[htbp]
  \centering
  \includegraphics[width=0.7\columnwidth]{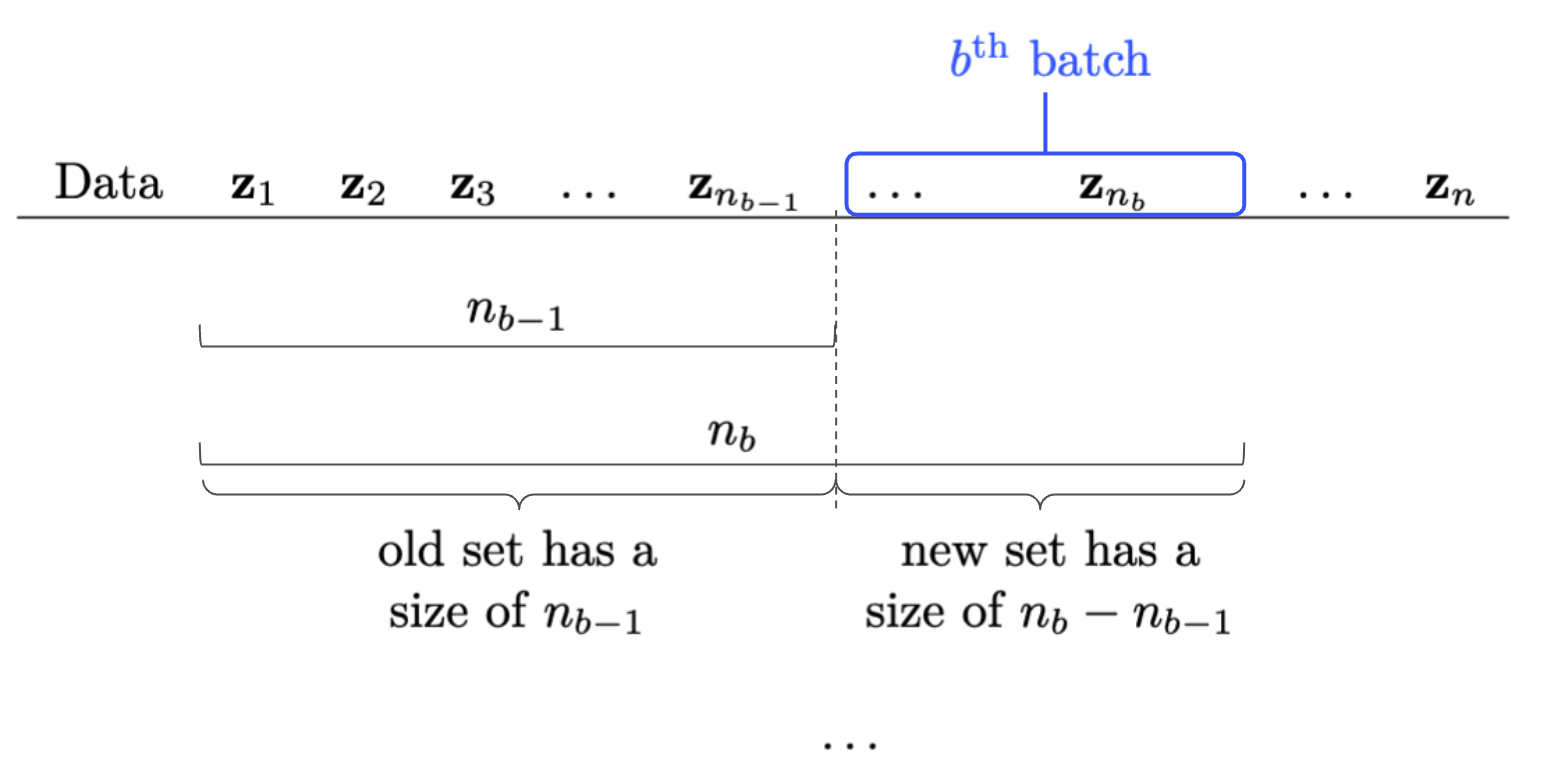}
  \caption{An illustration of the batch split.} \label{fig:adaptive_illustration}
\end{figure}

In general, we can consider splitting data into $B$ batches, where each batch $b$ (for $ b = 1, \ldots, B$) has a size $n_{b}-n_{b-1},$  with $n_0=0$. Figure \ref{fig:adaptive_illustration} illustrates the setup for this adaptive mechanism where the $b^{\text{th}}$ batch is observed after the posterior samples for the accumulated data of size $n_{b-1}$ have been obtained. Each batch is processed sequentially as outlined in Algorithm \ref{alg:anneal_SMC_adaptive}. Section \ref{sec:ex1_incremental} provides an example to illustrate the details with sequential data updates.

\begin{algorithm}[htbp]
\caption{\texttt{Adaptive\_Annealed\_SMC}}
\label{alg:anneal_SMC_adaptive}
\linespread{1.05}\selectfont
\small
\DontPrintSemicolon

\SetKwInOut{Input}{Input}
\SetKwInOut{Output}{Output}
\Input{
    (a) Data $\textbf{Z} = \left\{ \mathbf{z}_1, \ldots, \mathbf{z}_n\right\}$ and cumulative batch sizes $n_1, \ldots, n_B$; 
    (b) Dissimilarity metric $\mathcal{D}$; 
    (c) Tuning parameters $\phi, \epsilon$.
}
\Output{(a) Marginal likelihood estimates $\widehat{Z}_R$; (b) Posterior approximation, $\hat{\pi}(\particle) = \sum^K_{k=1} W_{R, k} \, \delta_{\particle_{R, k}}(\particle)$.}
\BlankLine
\For{$b \in \{1, 2, \ldots, B\}$}{
Calculate dissimilarities $\mathbf{D}$ from $\left\{ \mathbf{z}_1, \ldots, \mathbf{z}_{n_b}\right\}$ with a given dissimilarity metric $\mathcal{D}$.

\eIf{$b = 1$}{ \label{alg:initialize_s}
Set both $\mathbf{D}^{(0)}$ and $\tilde\pi_0\left(\cdot \right)$ to $\emptyset$.

Set $\tilde\pi_1\left(\textbf{x}_{i}\right)$ to $\mathcal{N}\left(\textbf{x}_{i}|\textbf{x}_{i}^{\text{CMDS}}, \rho \boldsymbol{I} \right)$ for $i=1, \ldots, n_1$.

}{

Set $\tilde\pi_0\left(\textbf{x}_i \right)$ to $\mathcal{N}\left(\textbf{x}_i|\hat{\textbf{x}}_i, \hat{\boldsymbol{\Sigma}}\right)$ for $i=1, \ldots, n_{b-1}$.

Set $\tilde\pi_1\left(\textbf{x}_{i}\right)$ to the prior distribution $\mathcal{N}\left(\textbf{x}_{i}|\textbf{0}, \Lambda\right)$ for $i=n_{b-1}+1, \ldots, n_{b}$.
}
       
$\{\particle_{0, k}\}_{k=1}^K \leftarrow$  \texttt{Particle\_Initialization} $\left(\mathbf{D}^{(0)}, \mathbf{D}, \tilde\pi_0, \tilde\pi_1, \pi(\parameter), K \right)$ \label{alg:initialize_e}

$\{(\particle_{R, k}, W_{R, k})\}_{k=1}^K, \widehat{Z}_R  \leftarrow$ \texttt{Annealed\_SMC} $\left( \{\particle_{0, k}\}_{k=1}^K, \tilde\pi_0, \tilde\pi_1, \pi(\parameter), L(\particle; \textbf{D}), \phi, \epsilon \right)$  \label{alg:posterior1}
 
Posterior approximation: $\hat{\pi}(\particle^{(b)}) = \sum^K_{k=1} W_{R, k} \, \delta_{\particle_{R, k}}(\particle)$. \label{alg:posterior2}

Compute weighted mean $\{\hat{\textbf{x}}_i\}_{i=1}^{n_b}$ and covariance  $\hat{\boldsymbol{\Sigma}}$ from $\{(\particle_{R, k}, W_{R, k})\}_{k=1}^K$. 

Reset $\mathbf{D}^{(0)} \leftarrow \mathbf{D}$.
}
\end{algorithm}

\section{Simulation Studies}
\label{sec:simulations}

The primary objective of the simulation studies is to evaluate the performance of various models under diverse data structures. We compared several candidate models, denoted as $\mathcal{M}_{g}^{\mathcal{D}}$, where $g$ represents the model distribution for dissimilarities and $\mathcal{D}$ is the dissimilarity metric used during estimation. We examined three experimental settings: one with known underlying dimensions, one with skewed errors, and the other with outliers. We also conducted a runtime analysis to compare the computation costs of various candidate models using both the proposed annealed SMC algorithm and the MCMC algorithm introduced in \citet{oh2001bayesian}. To maintain comparability between the two Bayesian methods, we kept a consistent computational budget. Initially, the annealed SMC was executed with $K$ particles, and the number of iterations was recorded. Subsequently, we equitably allocated this computational budget to MCMC by setting the number of MCMC iterations equal to the product of the annealed SMC iterations and the number of particles. For the implementation of the MCMC algorithm in \citet{oh2001bayesian}, we utilized the available R package BayMDS by \citet{oh2022baymds}, which implements the truncated Gaussian model. 

We established the values for the prior parameters by utilizing empirical Bayes methods, following the recommendations outlined in \citet{oh2001bayesian}. For the prior of $\sigma^2$, we chose $a = 5$ and $b = SSR/m$ obtained from CMDS. For the prior of $\psi$, we chose $c = -2$ and $d = 2$. For the hyperprior of $\lambda_k$, we set $\alpha = 1/2$ and  $\beta_k = \frac{1}{2}v_k^{(0)}/n$, where $v_k^{(0)}/n$ is the sample variance of the $k$th coordinate of $\textbf{X}$ from CMDS. For the mixing distribution of $\zeta_{i,j}$, we used degrees of freedom $\nu = 5$. The variance parameter for the reference distribution of the initial latent variables $\textbf{x}_{i}$ is set to $\rho = 0.01$. These parameter values deliver satisfactory results in the simulation studies, and the same values of the prior parameters are used in all examples unless otherwise specified.

In the random walk Metropolis-Hastings step, the constant multiplier of the variance of the Gaussian proposal density for generating $\textbf{x}_i$ and $\sigma^2$ is chosen based on the characteristics of the data to ensure rapid mixing.  In the annealed SMC algorithm, we set the number of particles to $K = 200$, the rCESS threshold to $\phi = 0.8$, and the resampling threshold to $\epsilon = 0.5$. Unless otherwise specified in the example, all parameters are sampled during the particle propagation steps. Parallel computation is also implemented for the annealed SMC algorithm with 3 cores across all simulations and examples.

\subsection{Experiment 1: Data with known true underlying dimension}

We start by investigating the model's performance in selecting the optimal dimension when the true underlying dimension is known within the data. The simulated data consists of 100 random samples of $\textbf{X}$ drawn from a 5-dimensional multivariate Gaussian distribution with mean 0 and variance $I$, the identity matrix. The true pairwise dissimilarities $\delta_{i,j}$ from $\textbf{X}$ are calculated as
\begin{center}
$\delta_{i,j} = \mathcal{D}(\mathbf{x}_i, \mathbf{x}_j), \qquad i,j = 1,\ldots,n,$
\end{center} 
where the dissimilarity metric $\mathcal{D}$ is chosen to be the Euclidean distance. Given $\delta_{i,j}$, the observed dissimilarities $d_{i,j}$ are generated from a normal distribution with mean $\delta_{i,j}$ and standard deviation of 1. These dissimilarities are truncated at 0. Thus, the simulated data consists of a 100 $\times$ 100 symmetric matrix of dissimilarities calculated from Euclidean distances with Gaussian errors. We included one frequentist method, CMDS, along with two Bayesian methods, MCMC and annealed SMC (ASMC), in the performance evaluation. In ASMC, all three candidate models ($\mathcal{M}_{TN}, \mathcal{M}_{TSN}$, and $\mathcal{M}_{TT}$) were considered, and both STRESS values and log marginal likelihood estimates (log $Z$) were reported. 
\begin{table}[htbp]

\ssmall
\begin{tabular}{|c|c|c|cccccc|}
\hline
\multirow{3}{*}{$p$} & \multirow{3}{*}{\begin{tabular}[c]{@{}c@{}}CMDS\\ STRESS\end{tabular}} & \multirow{3}{*}{\begin{tabular}[c]{@{}c@{}}MCMC \\ STRESS\end{tabular}} & \multicolumn{6}{c|}{ASMC} \\
 &  &  & \multicolumn{2}{c|}{$\mathcal{M}_{TN}^{\text{Euclidean}}$} & \multicolumn{2}{c|}{$\mathcal{M}_{TSN}^{\text{Euclidean}}$} & \multicolumn{2}{c|}{$\mathcal{M}_{TT}^{\text{Euclidean}}$} \\ 
 &  &  & \multicolumn{1}{c|}{STRESS} & \multicolumn{1}{c|}{logM} & \multicolumn{1}{c|}{STRESS} & \multicolumn{1}{c|}{logM} & \multicolumn{1}{c|}{STRESS} & logM \\ \hline
2 & 0.399 & 0.330 & \multicolumn{1}{c|}{0.396} & \multicolumn{1}{c|}{-5005.763} & \multicolumn{1}{c|}{0.397} & \multicolumn{1}{c|}{-7641.413} & \multicolumn{1}{c|}{0.351} & -7291.686 \\ \hline
3 & 0.266 & 0.268 & \multicolumn{1}{c|}{0.265} & \multicolumn{1}{c|}{-4869.795} & \multicolumn{1}{c|}{0.266} & \multicolumn{1}{c|}{-7454.384} & \multicolumn{1}{c|}{0.246} & -7178.706 \\ \hline
4 & 0.184 & 0.233 & \multicolumn{1}{c|}{0.186} & \multicolumn{1}{c|}{-4535.830} & \multicolumn{1}{c|}{0.188} & \multicolumn{1}{c|}{-7325.840} & \multicolumn{1}{c|}{0.187} & -7094.328 \\ \hline
5 & \textbf{0.163} & \textbf{0.223} & \multicolumn{1}{c|}{\textbf{0.168}} & \multicolumn{1}{c|}{\textbf{-4391.146}} & \multicolumn{1}{c|}{\textbf{0.172}} & \multicolumn{1}{c|}{\textbf{-7269.501}} & \multicolumn{1}{c|}{\textbf{0.171}} & \textbf{-7010.791} \\ \hline
6 & 0.199 & 0.232 & \multicolumn{1}{c|}{0.206} & \multicolumn{1}{c|}{-5058.168} & \multicolumn{1}{c|}{0.213} & \multicolumn{1}{c|}{-7803.021} & \multicolumn{1}{c|}{0.201} & -7612.594 \\ \hline
7 & 0.238 & 0.247 & \multicolumn{1}{c|}{0.240} & \multicolumn{1}{c|}{-5491.560} & \multicolumn{1}{c|}{0.246} & \multicolumn{1}{c|}{-7984.964} & \multicolumn{1}{c|}{0.243} & -7995.441 \\ \hline
8 & 0.278 & 0.262 & \multicolumn{1}{c|}{0.274} & \multicolumn{1}{c|}{-6033.348} & \multicolumn{1}{c|}{0.280} & \multicolumn{1}{c|}{-8409.323} & \multicolumn{1}{c|}{0.288} & -8689.978 \\ \hline
\end{tabular}
\caption{A summary of the STRESS values and log marginal likelihood estimates (logM) from applying the different MDS methods. The values in bold are the optimal dimensions selected by the lowest STRESS values or the largest log marginal likelihood estimates. All results are the averages of 20 runs. Bold values indicate the best performance for each case. \label{tab:sim1_tn_true}}

\end{table}

Table \ref{tab:sim1_tn_true} displays the STRESS values and log marginal likelihood estimates obtained from various MDS methods. All methods identified the true dimension as 5. Additionally, it can be observed that the candidate model $\mathcal{M}_{TN}^{\text{Euclidean}}$ attained the highest log marginal likelihood estimates and was considered the optimal model as expected.

\subsection{Experiment 2: Data with skewed errors}
\begin{figure}[htbp]
  \centering
  \subfloat[]{\includegraphics[width=0.35\columnwidth]{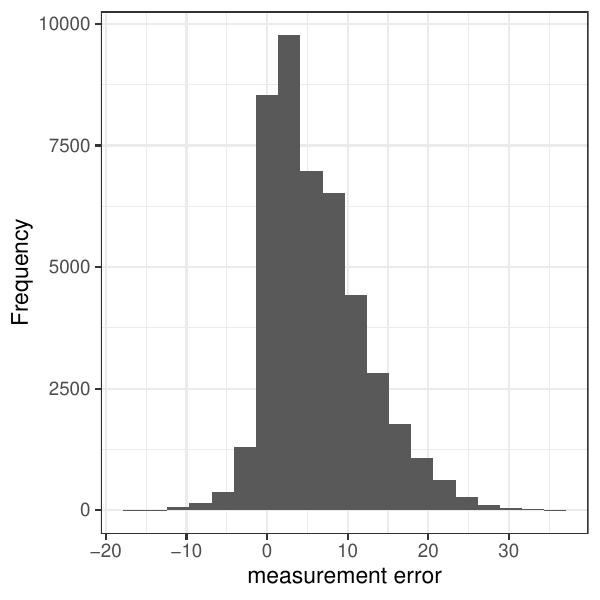}\label{fig:sim2_me_euc}}
  \subfloat[]{\includegraphics[width=0.65\columnwidth]{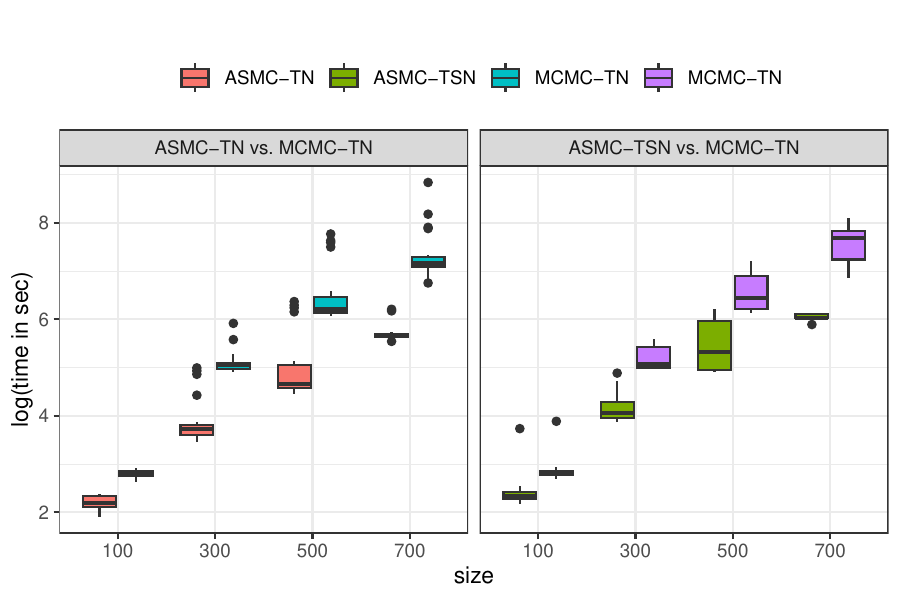}\label{fig:sim2_runtime}} \\
  \subfloat[]{\includegraphics[width=1\columnwidth]{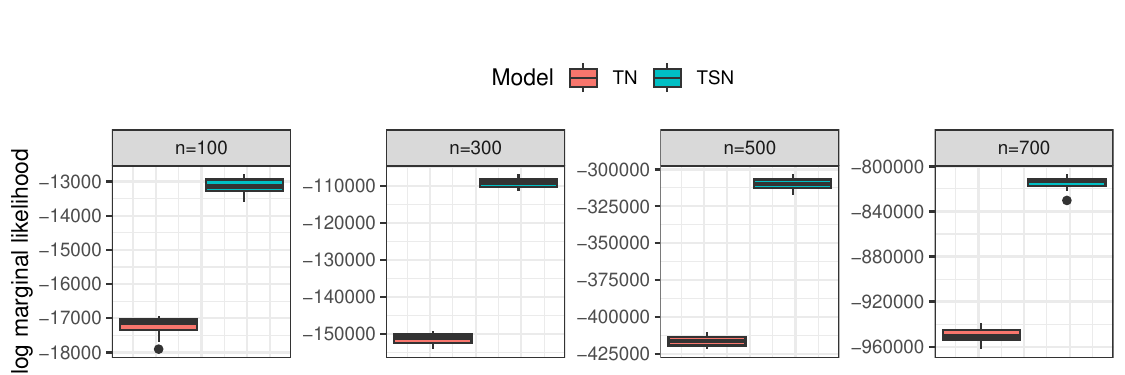}\label{fig:sim2_euclidean_logZ}}
  \caption{(a) The histogram of the errors with 300 observations. (b) The boxplots of computation times for different models. The computation budget is kept consistent across all comparisons between pairs of Bayesian methods. (c) The boxplots of the log marginal likelihood for different models.\label{fig:sim2_me_new}}
\end{figure}

In this example, our goal is to compare the performance of the proposed model $\mathcal{M}_{TSN}^{\text{Euclidean}}$ against the standard truncated Gaussian model $\mathcal{M}_{TN}^{\text{Euclidean}}$ in terms of log marginal likelihoods, particularly when the data are contaminated by skewed errors. We test how the proposed model performs when data skewness is present. We generated a dataset consisting of $n$ data points $\check{\textbf{Z}} = \left\{ \check{\mathbf{z}}_1, \ldots, \check{\mathbf{z}}_n\right\}$. Each of them has a dimension of 20, i.e., $\check{\mathbf{z}}_i = (\check{z}_{i,1}, \ldots, \check{z}_{i,20})^{\top}$. The data were simulated as a mixture of $50\%$ from $\mathcal{N}(0, 1)$, $25\%$ from $\mathcal{N}(100, 10)$ and $25\%$ from $\mathcal{N}(-10, 1)$. Next, we generated $n$ noisy observations $\textbf{Z}$ through a two-step process. First, we introduced minor errors into all data $\check{\textbf{Z}}$ to simulate systematic errors arising from data measurement. Second, varied percentages of the data points are subject to the contamination of moderate and significant errors, with the intention of replicating the scenario in which some observations are inaccurately recorded during data measurement. Specifically, moderate and significant errors were introduced into 20\% and 2\% of the data points, respectively. The Euclidean metric was then applied to obtain the dissimilarities $d_{i,j}$'s from the noisy data $\textbf{Z}$ and the dissimilarities $\check{d}_{i,j}$'s from clean data  $\check{\textbf{Z}}$. The errors $\epsilon_{i,j}$'s were computed by:
\begin{center}
    $\epsilon_{i,j} = d_{i,j} - \check{d}_{i,j},\qquad   i\neq j,\; i,j = 1,\ldots,n.$
\end{center} 
A detailed description of the data generation process with skewed errors is given in \textit{Supplementary}. We conducted experiments in several scenarios with sample sizes of $n=100, 300, 500, 700$ to assess the scalability of the model. Each scenario is repeated for 20 runs. A histogram of the errors $\epsilon_{i,j}$'s from one run with $n=300$ is shown in Figure \subref*{fig:sim2_me_euc}. 

Figure \subref*{fig:sim2_runtime} depicts the comparison of the computation times, showing the advantages of the annealed SMC algorithm over the MCMC algorithm across all scenarios with varying sample sizes. We examined four scenarios to compare the performance of the annealed SMC algorithm and the MCMC algorithm under the identical truncated Gaussian model. This comparison helped illustrate how computation time varies with increasing sample size. The computation times typically show greater variability as the sample size increases, as observed for $n=100, 300$, and $500$. 
These results illustrate that, even with more complicated models such as the truncated skewed Gaussian, our runtime remains comparable and even faster, demonstrating the scalability of the proposed framework to larger datasets. Figure \subref*{fig:sim2_euclidean_logZ} shows the performance comparison in terms of the log marginal likelihood as the skewed error presents. The findings of the study demonstrate that the model incorporating a truncated skewed Gaussian exhibits better performance, as the data under consideration is primarily with skewed errors. The results provide evidence that the skewed distributions are necessary for certain circumstances for modeling purposes.

\subsection{Experiment 3: Data with outliers}

\begin{figure}[htbp]
  \centering
  \subfloat[]{\includegraphics[scale=0.32]{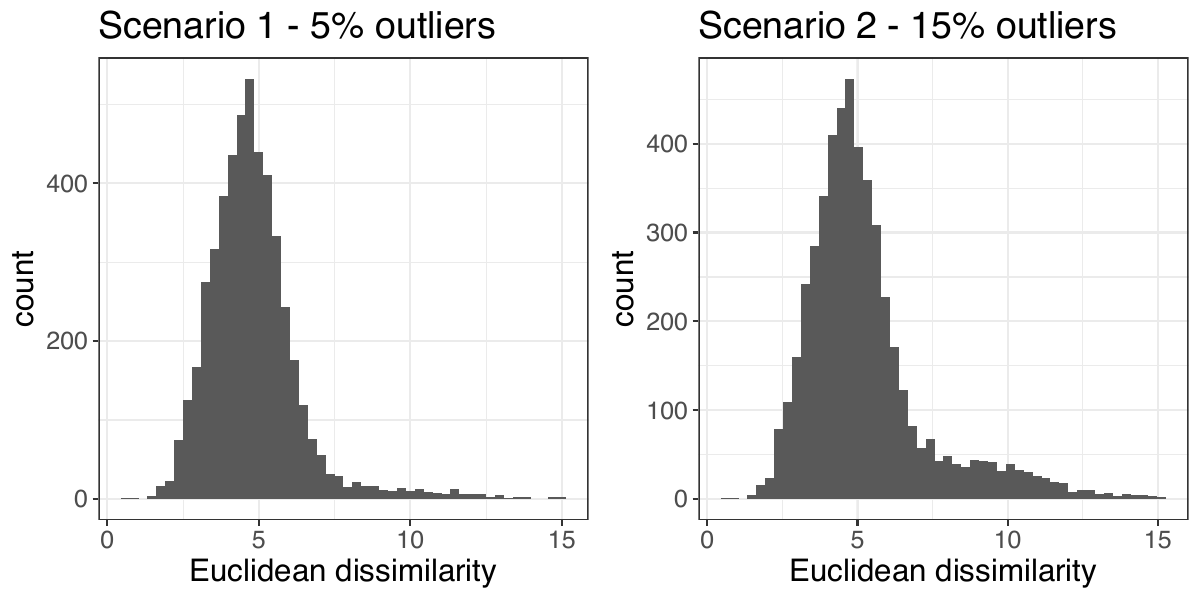}\label{fig:sim3_cont_d_hist}} 
  \subfloat[]{\includegraphics[scale=0.32]{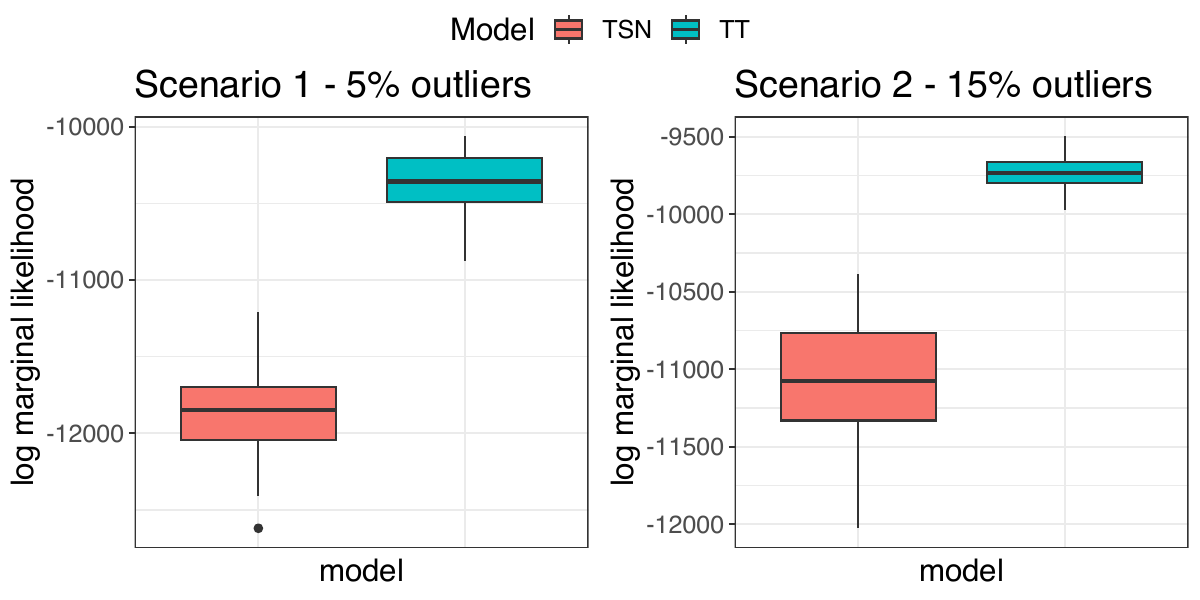}\label{fig:sim3_cont_logZ}}  \\
  \subfloat[]{\includegraphics[scale=0.50]{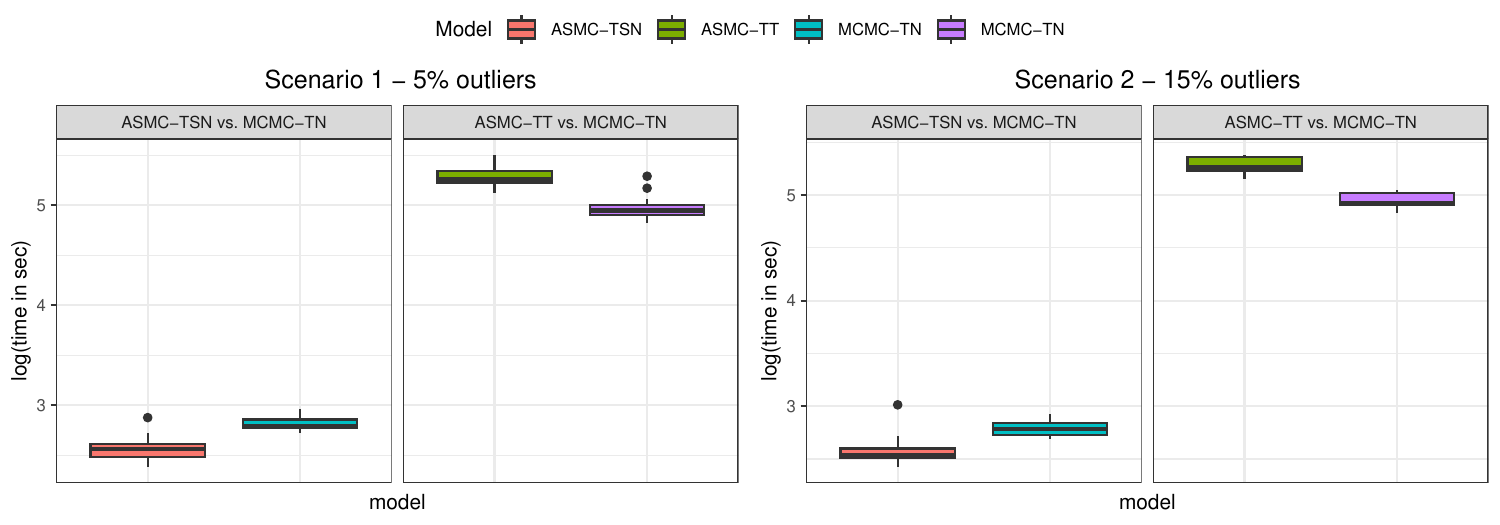}\label{fig:sim3_runtime}} 
  \caption{(a) The histograms of the dissimilarity $d_{i,j}$ under Euclidean metrics. The left histogram is from scenario 1, where the data contain 5\% outliers. The right histogram is from scenario 2, where the data contain 15\% outliers. (b) The boxplots of the log marginal likelihood for different models. Red: $\mathcal{M}_{TSN}^{\text{Euclidean}}$; Blue: $\mathcal{M}_{TT}^{\text{Euclidean}}$. Dimension $p$ is 2. (c) The boxplots of computation times for different models. The computation budget is kept consistent across all comparisons between pairs of Bayesian methods.}
\end{figure}

In the following experiment, we use the simulated data to investigate the robustness of the proposed models. The simulated dataset contains 100 observations which are generated from a 10-dimensional multivariate Gaussian distribution with mean 0 and variance $I$, the identity matrix. The observed dissimilarity matrix is generated by first computing the Euclidean dissimilarities from the raw observations and then adding some Gaussian errors centered around the true Euclidean dissimilarities, with a standard deviation of 0.5. To study the robustness of different models, we added outliers by randomly selecting a different proportion of the observed dissimilarities and doubling their values. We considered two scenarios with varying proportions of outliers in dissimilarities; the first contains 5\% outliers, and the second has 15\% outliers. Figure \subref*{fig:sim3_cont_d_hist} shows the histograms of the dissimilarity $d_{i,j}$'s under the two scenarios. In scenario 2, the increased percentage of outliers leads to a heavier tail in the dissimilarity histogram. In this simulation, we test the robustness of the models $\mathcal{M}_{TSN}^{\text{Euclidean}}$ and $\mathcal{M}_{TT}^{\text{Euclidean}}$.

Figure \subref*{fig:sim3_cont_logZ} shows the marginal likelihood in log scale for the two models under the two scenarios. In scenario 1, where the data only contained 5\% outliers, the histogram of the dissimilarity does not show an obvious heavy tail. According to the left boxplot in Figure \subref*{fig:sim3_cont_logZ}, the model $\mathcal{M}_{TT}^{\text{Euclidean}}$ is preferred since it produces higher log marginal likelihoods overall. When the percentage of outliers is increased to 15\%, an obviously heavier tail can be observed in the right histogram in Figure \subref*{fig:sim3_cont_d_hist}. The boxplot on the right in Figure \subref*{fig:sim3_cont_logZ} indicates that model $\mathcal{M}_{TT}^{\text{Euclidean}}$ is favored, as it yields higher log marginal likelihoods with smaller variability. As the percentage of outliers in the data increases, model $\mathcal{M}_{TSN}^{\text{Euclidean}}$ exhibits greater variance, while model $\mathcal{M}_{TT}^{\text{Euclidean}}$ demonstrates smaller variance across different seeds. Figure \subref*{fig:sim3_runtime} displays the comparison of the computation times between the annealed SMC algorithm and the MCMC algorithm across candidate models with different outlier percentages. It is evident that ASMC with $\mathcal{M}_{TSN}^{\text{Euclidean}}$ outperforms MCMC with $\mathcal{M}_{TN}^{\text{Euclidean}}$ in terms of computation time. When comparing ASMC with the more complex model $\mathcal{M}_{TT}^{\text{Euclidean}}$ to MCMC with the simpler model $\mathcal{M}_{TN}^{\text{Euclidean}}$, ASMC still demonstrates comparable runtime.

\section{Data Applications}
\label{sec:examples}

\subsection{NIH text data} \label{sec:ex3_text}

\begin{figure}[htbp]
  \centering 
  \subfloat[]{\includegraphics[scale=0.36]{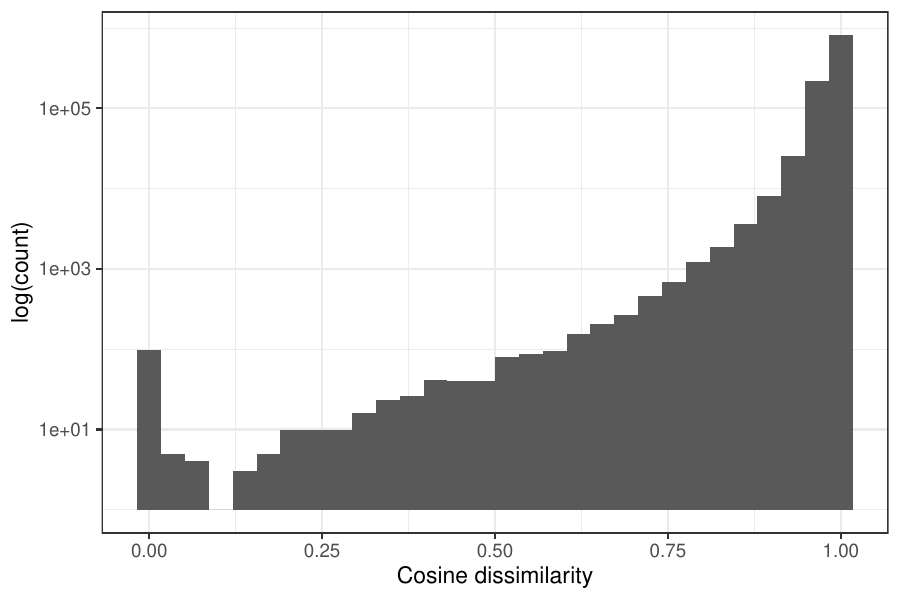}\label{fig:ex3_cosine_dis_log}}
  \subfloat[]{\includegraphics[scale=0.45]{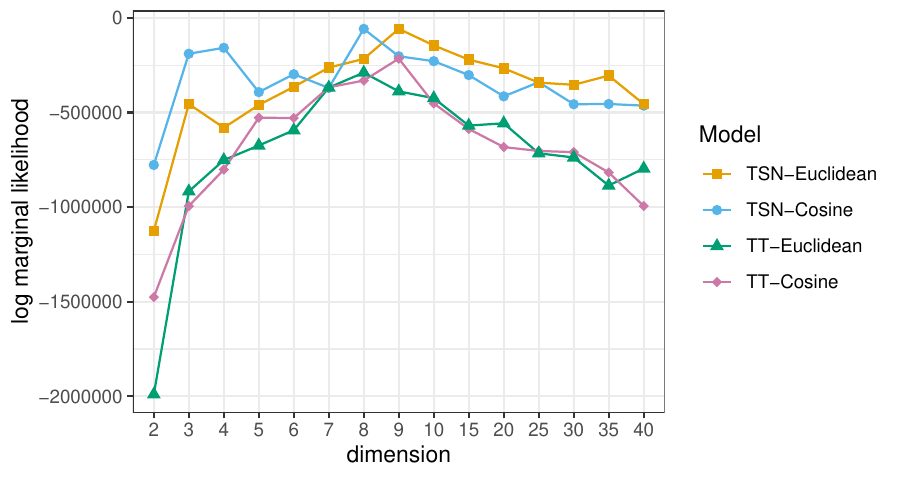}\label{fig:ex3_logZ}}
  \caption{(a) The histogram of Cosine dissimilarity. Counts on $Y$-axis are in log scale. (b) The log marginal likelihood for different models under varying dimensions.}
\end{figure}

In this example, we apply the MDS techniques to text data consisting of documents and words. We aim to showcase the application of our proposed method and investigate the effect of dimension $p$. The dataset holds information on research grants awarded by the National Institutes of Health (NIH) in 2022 \citep{NIHRePORTER}. The raw data contain 1471 grant abstracts from Saint Louis in the United States. To preprocess the data, we performed tokenization and removed stop words. This results in a document-term matrix with a dimension of 1471 by 238919, where each row represents an abstract and contains the word counts for all words in that abstract. We then re-weighted the word counts by multiplying an inverse document frequency vector to adjust for the relative importance of words in the entire collection of documents. The purpose of this reweighing step was to account for the varying frequencies of words across documents. We calculated the Cosine dissimilarities of the documents and used them as input to the MDS methods. Figure \ref{fig:ex3_cosine_dis_log} displays the histogram of Cosine dissimilarity values, with $Y$-axis on a log scale indicating the log count of occurrences. The histogram shows a spike at the lower end, around 0, and then a gradually increasing trend in counts as the Cosine dissimilarities approach 1. This distribution suggests that the dataset contains pairs with high similarity (values near 0) and high dissimilarity (values near 1), with a notable concentration at higher dissimilarity levels.

We varied the dimension $p$ across values from 2 to 10, as well as 15, 20, 25, 30, 35, and 40, to examine its impact on the results. All four candidate models are included in the comparison. Figure \ref{fig:ex3_logZ} presents the relationship between model dimensionality and log marginal likelihood estimates for four different models, represented by distinct colored markers. Each line represents a different model's performance across dimensions, with log marginal likelihood estimates initially increasing as the dimension rises, reaching a peak around dimensions 8 to 10 for most models, and then slightly decreasing as dimensionality continues to increase. In general, models with higher log marginal likelihood estimates are considered to perform better. In this case, model $\mathcal{M}^{\text{Euclidean}}_{TSN}$ reaches the highest overall log marginal likelihood estimates at dimension $p=9$, indicating superior performance in higher dimensions. However, model $\mathcal{M}^{\text{Cosine}}_{TSN}$ exhibits a notable advantage in lower dimensions, achieving relatively high log marginal likelihood estimates when $p$ is small. This suggests that while $\mathcal{M}^{\text{Euclidean}}_{TSN}$ is optimal in higher dimensions, $\mathcal{M}^{\text{Cosine}}_{TSN}$ may be more suitable or efficient when the dimensionality is low. Thus, there may be trade-offs between models based on the dimension, with $\mathcal{M}^{\text{Cosine}}_{TSN}$ performing better at lower dimensions and $\mathcal{M}^{\text{Euclidean}}_{TSN}$ excelling as the dimension increases.

\subsection{Clustering abstract text data} \label{sec:ex4_text}

This example explores an interesting application of MDS, clustering objects by grouping similar observations based on their dissimilarities. With MDS techniques, the coordinates of objects can be represented in a low-dimensional space. Visualizing these clusters in the reduced-dimensional space is of interest, as it can provide meaningful insights into the relationships between the groups. The dataset comprises 139 English abstracts collected from three academic journals during 2022 and 2023: 44 from \textit{Annals of Mathematics}, 50 from \textit{Bayesian Analysis}, and 45 from \textit{Biometrics}. While all three journals focus on the development of statistical and mathematical methods, each emphasizes distinct application areas. This study aims to assess the performance of the proposed model in clustering these abstract texts. We employed the Jaccard metric with a 2-gram model to construct the dissimilarity matrix. A 2-gram model represents a subsequence of two successive elements within an abstract, including words, numbers, or symbols. Each abstract was first treated as a set of unique 2-grams, and the Jaccard dissimilarity was computed based on the intersection and union of these sets. This metric is effective for identifying abstracts with overlapping vocabularies or patterns, mainly when abstracts are dense with technical terms. Additionally, the use of a 2-gram model is beneficial in applications where sequences of words are relevant.

\begin{figure}[htbp]
  \centering
  {\includegraphics[scale=0.8]{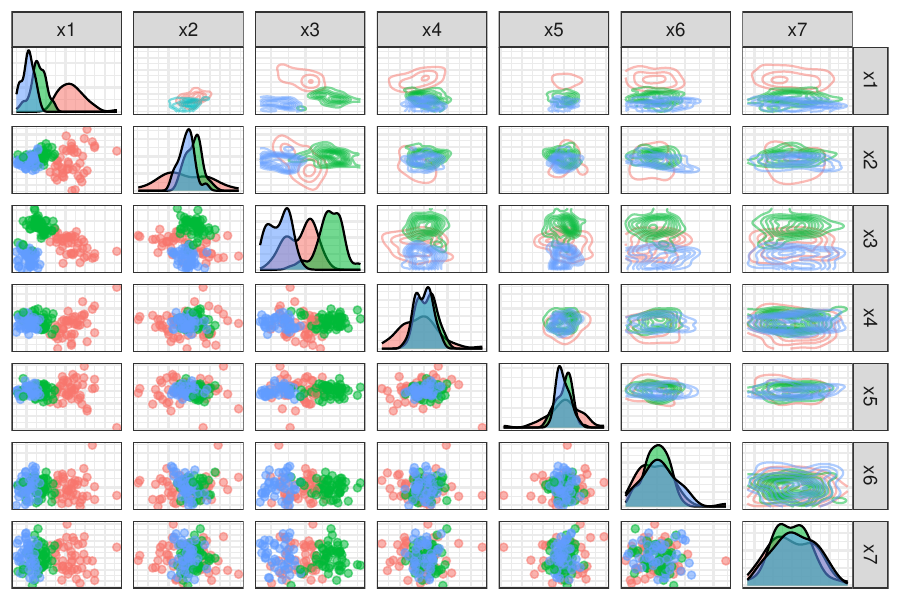}} 
  \caption{Estimated coordinates of the abstracts obtained using GBMDS-ASMC with $p=7$. The lower panel displays pairwise scatter plots, the diagonal shows density plots, and the upper panel provides contour plots of the density. The colors indicate three distinct clusters.} \label{fig:ex4_cluster}
\end{figure}

We applied GBMDS-ASMC for reducing dimension, followed by the partition around medoids algorithm (PAM) \citep{kaufman1990partitioning} to group the results into 3 clusters. In this study, we predetermined the number of clusters as 3, corresponding to the three journals included in the dataset. All candidate models were evaluated, and the model $\mathcal{M}^{\text{Euclidean}}_{TSN}$ at dimension $p=7$ achieved the highest log marginal likelihood estimates. Therefore, it was selected as the optimal model for this example. Figure \ref{fig:ex4_cluster} presents the pairwise scatterplots and density plots of the clustering results, highlighting that the first three dimensions contribute significantly to the clustering, with clear separations between clusters. This result provides a meaningful representation of the data, yielding well-defined clusters with visual separation. We also evaluated the classification accuracy as another performance metric using the known journal labels of each abstract. The GBMDS-ASMC method correctly classified all abstracts into their respective journals. Table \ref{tab:ex4_topwords} summarizes the top 20 keywords from each cluster along with their corresponding journals. The results reveal the presence of similar keywords related to the development of statistical and mathematical methods. Additionally, distinct keywords are observed, reflecting the specific application and thematic focus unique to each journal.

\begin{table}[htbp]
\small
\begin{tabular}{|c|c|c|}
\hline
\textbf{Cluster} & \textbf{Keywords} & \textbf{Journal} \\ \hline
1 & \begin{tabular}[c]{@{}c@{}}conjecture, prove, theory, smooth, \\ groups, minimal, stable, Euler, symmetric,  \\ finite, hypersurfaces, singularity, \\ characteristic, spaces, field, cohomology, \\ Frobenius, algebra, existence, invariant\end{tabular} & \textit{Annals of Mathematics} \\ \hline
2 & \begin{tabular}[c]{@{}c@{}}Bayesian, model, posterior, prior, Gaussian,\\ distribution, inference, Monte-Carlo, spatial, \\ network, linear, MCMC, hierarchical, mixture, \\ predictive, sampling, computational, \\ Dirichlet-process, shrinkage, uncertainty\end{tabular} & \textit{Bayesian Analysis} \\ \hline
3 & \begin{tabular}[c]{@{}c@{}}estimator, treatment, data, effect, \\ regression, causal, randomized, covariate, \\ functional, dose, population, time, \\ consistent, hazards, heterogeneity, confounding, \\ asymptotic, trials, Cox, longitudinal\end{tabular} & \textit{Biometrics} \\ \hline
\end{tabular}
\caption{A summary of the top 20 keywords from each cluster on the abstract data. \label{tab:ex4_topwords}}
\end{table}

\subsection{NeurIPS
 text data with incremental data updates} \label{sec:ex1_incremental}

In the first example, we demonstrate the performance of the adaptive inference with the annealed SMC algorithm on text data with incremental size. The text data is generated from articles from the Conference on Neural Information Processing Systems (NeurIPS). The NeurIPS dataset contains NeurIPS conference papers published between 1987 and 2015 \citep{Dua:2019, perrone2017poisson}. In this study, our focus is directed toward a subset of the NeurIPS dataset, comprising a matrix of word counts extracted from 55 articles. This matrix is referred to as the document-term matrix, which is constructed after tokenization, stop-word removal, and truncation of the corpus by retaining only words that appear more than fifty times. The document-term matrix has counts for a list of 15005 words. Instead of Euclidean dissimilarity, we consider Cosine dissimilarity, which is suitable for discrete data such as word counts since it measures how dissimilar the documents are irrespective of their sizes. 

We fitted the model $\mathcal{M}^{\text{Cosine}}_{TT}$ and compared the results by STRESS values and computational times. The number of significant attributes $p$ is assumed to be 2. In this toy example, we considered the cases with $m_0 = 10, 50$, and $m_1 = 1, 5$. For each combination of $m_0$ and $m_1$, we considered two cases:  one applies the annealed SMC algorithm to the combined $n=m_0+m_1$  observations from scratch, while the other employs the annealed SMC algorithm with sequential data updates, leveraging the 
posterior samples obtained from the first  $m_0$ observations. We set the number of particles to $K = 100$ and the rCESS threshold to $\phi = 0.9$. At each particle propagation step,  a subset of two parameters is randomly selected for updating. If the latent variables $\left\{ \mathbf{x}_1, \ldots, \mathbf{x}_n\right\}$ are selected, 
 a random subset containing half of the components is updated. This setting balances computational efficiency with effective exploration of the parameter space.

From Table \ref{tab:incr_ASMC_res}, it can be seen that the STRESS values from both cases are close, and this validates the performance of the annealed SMC algorithm for incremental dimension. For the two cases with $m_0 = 10$, the computation times decrease by an average of 24\%  when applying the annealed SMC algorithm for incremental data. Similarly, for the two cases with $m_0 = 50$, the computation times fall by 31\%. Moreover, the variability in computation times decreased in both cases with the implementation of adaptive inference.

\begin{table}[htbp]
\scriptsize
\begin{tabular}{c|ccllc|cc}
Observations     & STRESS          & \begin{tabular}[c]{@{}c@{}}Time \\ (in sec)\end{tabular} & \multicolumn{1}{c}{} & \multicolumn{1}{c}{} & Observations     & STRESS & \begin{tabular}[c]{@{}c@{}}Time\\ (in sec)\end{tabular} \\  
n = 11           & 0.7093          & 19.0 (4.2)                                                    &                      &                      & n = 51           & 0.7449 & 143.0 (36.3)                                                  \\
$m_0 = 10, m_1 = 1$ & 0.7133          & 15.1 (3.1)                                                  &                      &                      & $m_0 = 50, m_1 = 1$ & 0.7419 & 100.9 (13.9)                                                  \\
n = 15           & 0.7194         & 29.4 (9.9)                                                    &                      &                      & n = 55           & 0.7545 & 156.5 (37.0)                                                  \\
$m_0 = 10, m_1 = 5$ & 0.6930 & 21.6 (4.0)                                                   &                      &                      & $m_0 = 50, m_1 = 5$ & 0.7536 & 104.4 (14.4)                                               
\end{tabular}
\caption{A summary of the STRESS values and computation times from applying the annealed SMC on GBMDS to observations with different dimensions. The first and third rows present the results from applying the annealed SMC algorithm of fixed dimension to all observations. The results in the second and fourth rows come from running the annealed SMC algorithm of incremental dimension, given the results from $n$ observations are known. The values in the parentheses are the standard deviation of the computation times. All results are the averages of 30 runs. \label{tab:incr_ASMC_res}}
\end{table}

\subsection{Geographical data} \label{sec:ex2_geographical}
This example aims to study the performance of the proposed method and visualize the estimations with uncertainty measures. As an illustrative example to compare CMDS and BMDS, we considered the US cities dataset from the US Census Bureau \citep{UScensus}, which contains Latitude and Longitude information for 15 large US cities. To evaluate the robustness of GBMDS, we appended 10 noise variables generated from a Gaussian distribution with a mean of 0 and variance comparable to the ``Latitude'' or ``Longitude'' variables.

 We performed experiments across three scenarios, differing in the weight assigned to true signal versus noise. We define the signal-to-noise ratio as $R_\text{s:n}$ as the ratio of the total weight of the true signal variables to that of the noise. In Scenario 1 ($R_\text{s:n}=1$), the total weight of the two signal variables equals the total weight of the 10 noise variables. In Scenario 2 ($R_\text{s:n}=4$), emphasis is placed on the geographic variables. Scenario 3 ($R_\text{s:n}=10$) represents an extreme condition where the majority of the weight is allocated to the signal. In all experiments, weights were normalized to sum to 1.
 
\begin{table}[htbp]
\small
\caption{A Summary of the STRESS values for different methods on US City data under scenarios 1 to 3 with different noise-to-signal ratios. Bold values indicate the best performance for each case.  \label{tab:STRESS}}
\begin{tabular}{lcc}
                                            & CMDS    & GBMDS-ASMC \\\hline
Scenario 1: $R_\text{s:n}=1$       & 0.4557      & \textbf{0.3521} \\
Scenario 2: $R_\text{s:n}=4$       & 0.4680      & \textbf{0.3910} \\
Scenario 3: $R_\text{s:n}=10$      & 0.4726      & \textbf{0.4103} \\
\end{tabular}
\end{table}

To implement the proposed GBMDS, we utilized the annealed SMC (ASMC) algorithm, initialized using results from CMDS. For simplicity, we assumed $\psi = 0$, reducing the model to $\mathcal{M}_{TN}^{\text{Euclidean}}$. While standard MCMC methods yield similar results for this small dataset1, we report only the ASMC results to maintain focus on the comparison between Bayesian and frequentist approaches.

\begin{figure}[htbp]
  \centering
  \subfloat[]{\includegraphics[width=0.33\columnwidth]{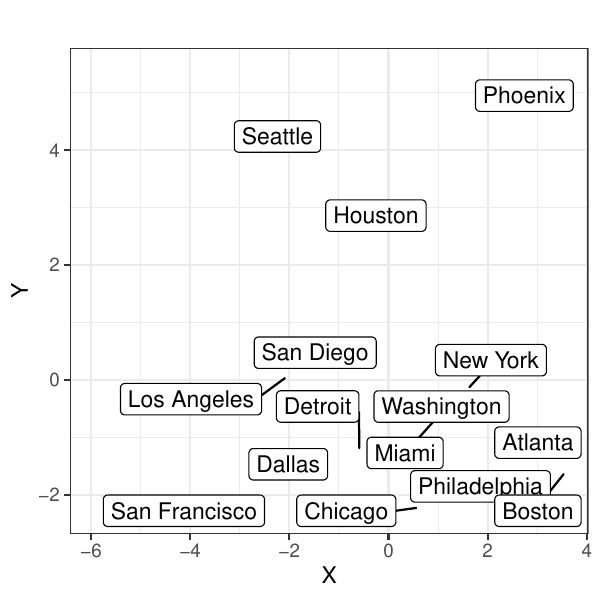}\label{fig:ex2_cmds_R_1}}
  \subfloat[]{\includegraphics[width=0.33\columnwidth]{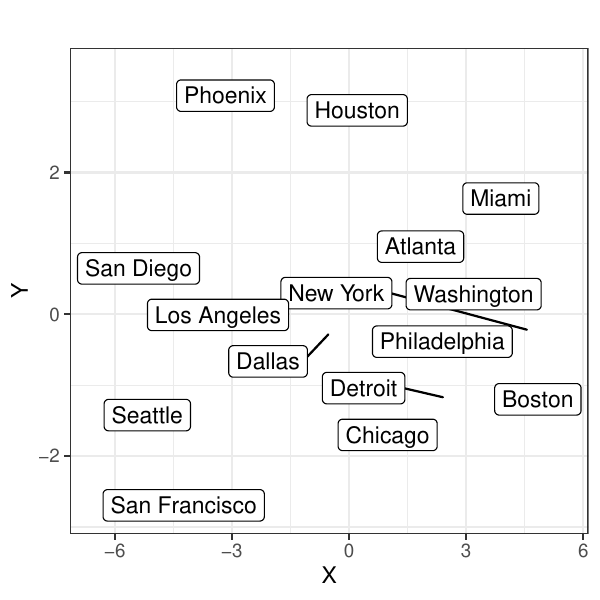}\label{fig:ex2_cmds_R_4}} 
  \subfloat[]{\includegraphics[width=0.33\columnwidth]{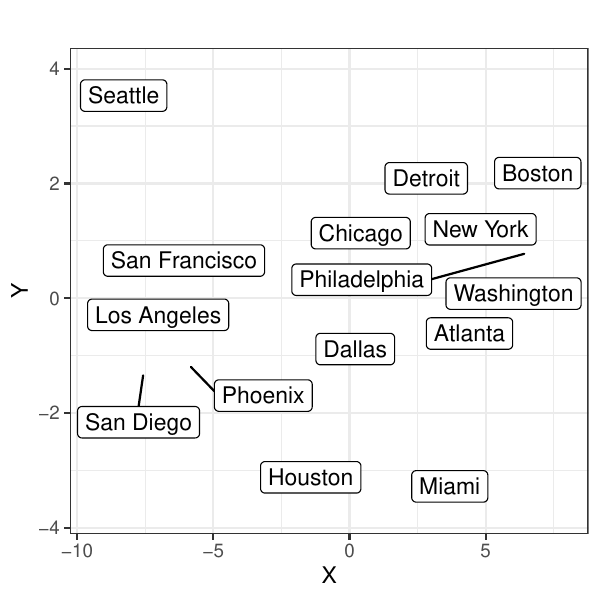}\label{fig:ex2_cmds_R_10}}
  \par\medskip
  \subfloat[]{\includegraphics[width=0.33\columnwidth]{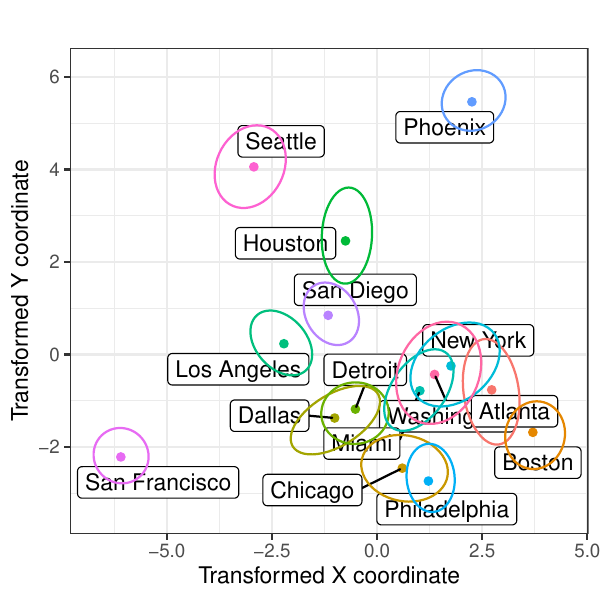}\label{fig:ex2_bmds_R_1}}
  \subfloat[]{\includegraphics[width=0.33\columnwidth]{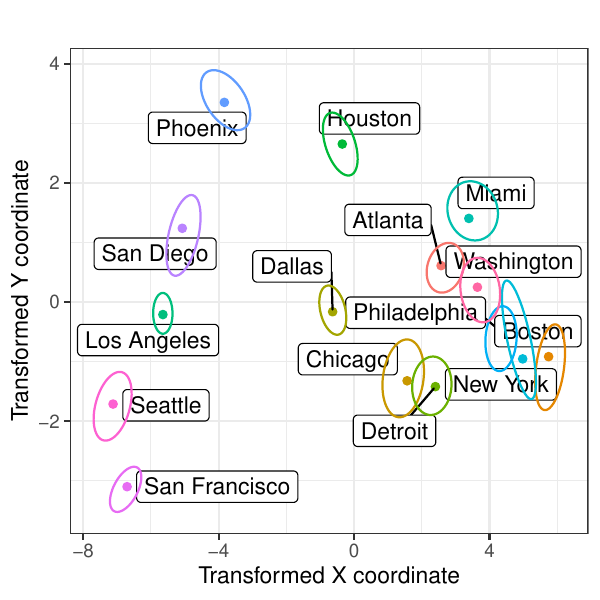}\label{fig:ex2_bmds_R_4}} 
  \subfloat[]{\includegraphics[width=0.33\columnwidth]{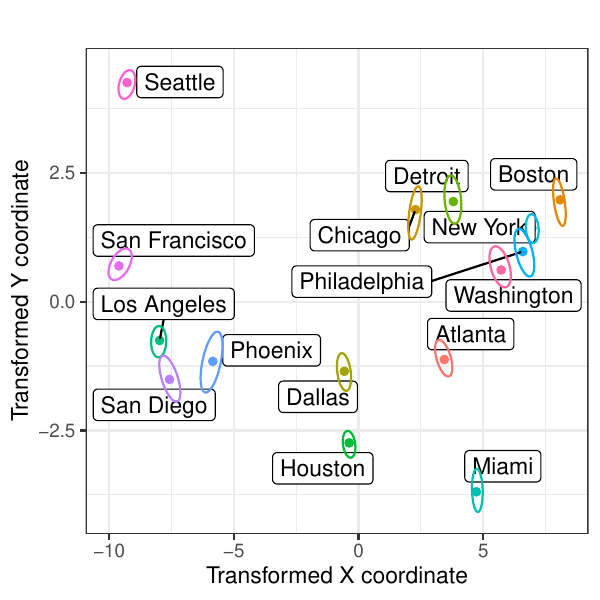}\label{fig:ex2_bmds_R_10}}
  \caption{Estimated locations of the 15 US cities from CMDS and GBMDS-ASMC after transformations. Sub-figures (a) to (c) are results from CMDS and (d) to (f) are from GBMDS-ASMC. $R_\text{s:n}=1$ in (a) and (d), $R_\text{s:n}=4$ in (b) and (e), $R_\text{s:n}=10$ in (c) and (f). For GBMDS-ASMC, the ellipses are generated from all the posterior samples with the 95\% credible regions. The posterior medians of $\textbf{x}_i$'s are served as the estimated coordinates of the 15 US cities in the two-dimensional space. \label{fig:ex2_ASMC}}
\end{figure}

Table \ref{tab:STRESS} presents the STRESS values obtained from CMDS and GBMDS (via ASMC) for the three scenarios. The Bayesian approach yields lower STRESS values across all cases. Figure \ref{fig:ex2_ASMC} displays the estimated locations of the 15 US cities. Procrustes transformations (rotation and reflection) were applied to align the estimates with the actual map. As observed in Figures \subref*{fig:ex2_cmds_R_1} and \subref*{fig:ex2_bmds_R_1}, under the equal weight scenario ($R_\text{s:n}=1$), several cities are geographically misplaced regardless of transformation, as the geographic signal is masked by the noise variables. However, the estimated locations shown in \subref*{fig:ex2_cmds_R_10} and \subref*{fig:ex2_bmds_R_10} show a much closer match when higher weights are assigned to the ``Latitude'' and ``Longitude'' variables.

The Bayesian approach offers several advantages over the classical approach. Beyond producing lower STRESS values, the Bayesian approach enables the estimation of uncertainty by leveraging samples from the posterior distribution. Since MDS solutions are invariant to rotation and translation,  we performed Procrustes transformations on the posterior samples of $\textbf{x}_i$ to align them to a reference configuration (the posterior mean estimate), effectively standardizing the orientation of all samples of $\textbf{x}_i$. Using these aligned samples, we constructed credible regions, represented as ellipses in Figures \subref*{fig:ex2_bmds_R_1} to \subref*{fig:ex2_bmds_R_10}. In contrast to CMDS, our GBMDS-ASMC method provides quantifiable uncertainty measures, exhibiting tight credible regions in scenarios with high signal-to-noise ratios 
and wider regions when the data contains significant noise.

\section{Conclusion and Discussion}
\label{sec:conclusion}

In this work, we introduced a generalized Bayesian multidimensional scaling (GBMDS) framework that addresses the limitations of existing MCMC-based BMDS methods. By incorporating non-Gaussian error distributions and diverse dissimilarity metrics, we enhanced the robustness and flexibility of MDS. We also developed an adaptive ASMC algorithm for efficient Bayesian inference. Unlike standard MCMC,  which requires separate, often complex procedures to estimate model evidence, our algorithm provides a marginal likelihood estimator naturally as a byproduct of sampling. This facilitates rigorous model comparison across candidate models with varying error distributions, dissimilarity metrics, and dimensions.

Bayesian model selection offers distinct advantages over classical approaches such as $p$-values and information criteria. First, Bayes factors provide a direct probabilistic measure of model support, offering  intuitive interpretation. Second, Bayesian model selection is consistent, meaning that given sufficient data, it  asymptotically selects the model minimizing the Kullback-Leibler divergence to the true model under mild conditions \citep{berk1966limiting}. Third, it  naturally penalizes excessive model complexity,  favoring parsimony\citep{berger2001objective,robert2007bayesian}.

Leveraging these advantages, we propose selecting the optimal model-dimension combination for GBMDS using Bayes factors. While frequentist model selection via STRESS relies on loss statistics tailored to particular cases, our fully Bayesian approach utilizes the marginal likelihood estimates from the adaptive ASMC algorithm. Although adaptive schemes may introduce negligible bias, they provide significant gains in robustness and variance reduction compared to fixed schedules. This framework eliminates the need for additional computational steps, offering a principled and efficient alternative to existing MCMC-based estimators.

The proposed annealed SMC algorithm offers a robust alternative to MCMC, particularly when computational resources permit parallelization or when robust inference is prioritized over single-chain MCMC exploration. The adaptive ASMC framework is highly flexible, allowing for the straightforward integration of advanced MCMC kernels from existing literature, such as Hamiltonian Monte Carlo (HMC), to further enhance mixing efficiency. Regarding memory constraints, while high-dimensional particle systems can present challenges, these are mitigated in BMDS by the typically low dimensionality of the latent embeddings. Furthermore, our method eliminates the need to store the complete particle history at every annealing step, significantly reducing overhead; consequently, even standard personal computing hardware can efficiently handle substantial particle populations. Should memory limitations arise in large-scale applications, they can be effectively managed by trading particle count for annealing iterations, which is a strategy shown to maintain or improve performance with appropriate tuning \citep{wang2020annealed}. Alternatively, scalability can be ensured through distributed computing strategies such as entangled Monte Carlo \citep{NIPS2012_dc4c44f6}, the island particle filter \citep{Verge_island_smc_2015}, and the distributed particle filter \citep{HEINE20172508}, ensuring that memory limitations do not hinder the application of SMC to large datasets

The proposed method is implemented in the R package \texttt{GBMDS} (\url{https://github.com/SFU-Stat-ML/GBMDS}), integrating  R and C++ for efficiency. 
Comparison with the MCMC-based package \texttt{BayMDS}  \citep{oh2022baymds} demonstrated that our method achieves comparable or faster runtimes even for complex models (e.g., truncated Student's $t$), and successfully handles datasets exceeding 1,000 observations where traditional MCMC implementations faced capacity limitations.

Despite the methodological advancements of the proposed GBMDS framework, several limitations remain. First, addressing the inherent non-identifiability of the latent configuration relies on post-hoc Procrustes transformations. While this alignment allows for the construction of credible regions, we observed that these regions can remain relatively broad in scenarios with high noise or weak signal, potentially obscuring fine-grained spatial patterns. 
Second, although our adaptive algorithm significantly improves computational efficiency over MCMC, scalability to massive datasets remains a challenge. Specifically, the memory required to store and propagate the particle population grows linearly with both the sample size and the latent dimension. While our batch processing strategy mitigates this, extremely large datasets may still necessitate further approximation techniques or distributed computing implementations to remain computationally feasible. Third, the proposed adaptive algorithm currently assumes a fixed global latent dimension $p$ across all data batches. This assumption may be restrictive in streaming or sequential settings where the intrinsic dimensionality of the data varies over time or across partitions, necessitating more complex alignment or dimension-jumping mechanisms.

There are several directions for future work. First, while we utilized Procrustes transformations to address non-identifiability and visualize uncertainty, exploring alternative post-processing transformations could further enhance interpretability. Second, our current framework assumes a fixed latent dimension; future work should address cases where intrinsic dimensionality varies across data batches or partitions, potentially requiring alignment steps for valid comparison. Third, extending the adaptive ASMC algorithm to perform joint MDS and clustering could further enhance the versatility of the framework in detecting structure. Finally, future research should focus on enhancing the computational efficiency of the algorithm. Key directions include integrating established MCMC kernels from the recent BMDS literature, adopting subsampling-based inference \citep{gunawan2020subsampling} to reduce computational costs, and implementing advanced parallelization strategies to improve scalability for increasingly large datasets.

\backmatter

\bmhead{Supplementary information}

Supplementary materials include a document with the details of the dissimilarity metrics, details of the particle propagation step, and details of the data-generating process in simulation studies.

\bibliography{bibliography}

\end{document}